\documentclass[sigconf,review,10pt]{acmart}
\AtBeginDocument{%
  }

\setcopyright{none}
\copyrightyear{2018}
\acmYear{2018}
\usepackage{xspace}
\usepackage{enumitem}
\usepackage{array}
\usepackage{tabularx}
\usepackage{makecell}
\usepackage{multirow}
\usepackage[normalem]{ulem}
\usepackage{url}
\newcommand{\name}{{VeriCache}\xspace}

\usepackage{amsthm}
\usepackage{algorithm}
\usepackage{algpseudocode}
\usepackage{listings}

\lstdefinestyle{vcpython}{%
  language=Python,
  basicstyle=\footnotesize\ttfamily,
  keywordstyle=\color{black}\bfseries,
  commentstyle=\color{gray}\itshape,
  showstringspaces=false,
  columns=fullflexible,
  keepspaces=true,
  aboveskip=4pt,
  belowskip=4pt,
}
\newtheorem{insight}{Insight}

\setlist{topsep=1pt,partopsep=0pt,parsep=0pt,itemsep=1pt}

\makeatletter
\renewcommand\section{\def\@toclevel{1}%
  \@startsection{section}{1}{\z@}%
  {-0.3\baselineskip \@plus -0.5\p@ \@minus -0.2\p@}%
  {0.1\baselineskip}%
  {\ACM@NRadjust\@secfont}}
\renewcommand\subsection{\def\@toclevel{2}%
  \@startsection{subsection}{2}{\z@}%
  {-0.25\baselineskip \@plus -0.5\p@ \@minus -0.2\p@}%
  {0.1\baselineskip}%
  {\ACM@NRadjust\@subsecfont}}
\renewcommand\subsubsection{\def\@toclevel{3}%
  \@startsection{subsubsection}{3}{\z@}%
  {-0.2\baselineskip \@plus -0.5\p@ \@minus -0.2\p@}%
  {-3.5\p@}%
  {\ACM@NRadjust{\@subsubsecfont\@adddotafter}}}
\renewcommand\paragraph{\def\@toclevel{4}%
  \@startsection{paragraph}{4}{\parindent}%
  {1\p@ \@plus 0.3\p@ \@minus 0.2\p@}%
  {-0.5em}%
  {\ACM@NRadjust{\@parfont\@adddotafter}}}
\let\ACM@origsection\section
\let\ACM@origsubsection\subsection
\let\ACM@origsubsubsection\subsubsection
\let\ACM@origparagraph\paragraph
\makeatother

\begin{document}

\title{\name: Turning Lossy KV Cache into Lossless LLM Inference}
%\title{\name: Lossless LLM Inference over Compressed KV Caches}
\author{\normalsize Jiayi Yao\textsuperscript{1}, Samuel Shen\textsuperscript{2}, Kuntai Du\textsuperscript{2}, Shaoting Feng\textsuperscript{1}, Dongjoo Seo\textsuperscript{3}, Rui Zhang,\\ Yuyang Huang\textsuperscript{1}, Yuhan Liu\textsuperscript{1}, Shan Lu\textsuperscript{4,1}, Junchen Jiang\textsuperscript{2,1}}
\affiliation{%
  \institution{\textsuperscript{1}University of Chicago \quad \textsuperscript{2}Tensormesh Inc. \quad \textsuperscript{3}Samsung Semiconductor \quad \textsuperscript{4}Microsoft Research}
  \country{}
}
\email{}

\settopmatter{printacmref=false}
\renewcommand\footnotetextcopyrightpermission[1]{}
\pagestyle{plain}

\renewcommand{\shortauthors}{Yao et al.}


\settopmatter{printacmref=false}

\received{20 February 2007}
\received[revised]{12 March 2009}
\received[accepted]{5 June 2009}

\begin{abstract}
The large size of the KV cache has become a major bottleneck for serving LLMs with increasing context lengths. In response, many KV cache compression methods, such as token dropping and quantization, have been proposed. However, almost all of these methods are inherently lossy---despite minimal accuracy degradation for short outputs, their outputs increasingly diverge from full-KV-cache outputs as more tokens are decoded, which leads to catastrophic failures in code generation and tool calling.

We present \name, the first inference framework that ensures the {\em same} output as full-KV-cache decoding but largely preserves the high decoding throughput of a \textit{range} of KV cache compression algorithms.
\name uses the compressed KV cache to draft tokens, then verifies them against the full KV cache.
While it may seem like just speculative decoding, \name requires addressing a key system challenge to work---keeping the full KV cache out of GPU memory {\em and} minimizing the overhead of swapping it in for verification.
The insight is two-fold:
(1) compressed-KV decoding can be parallelized with full-KV swap, because one is HBM-bandwidth-bound and the other is PCIe/network-bound, and
(2) the compressed KV cache often produces output similar to the full KV cache, allowing a long drafting horizon to amortize each full-KV swap.

\name applies to both long-context decoding and remote prefix caching, supports a broad family of token-dropping and quantization methods through a uniform compressor interface, and composes with traditional speculative decoding.
Experimental results show that \name achieves up to $4\times$ higher throughput than full-KV inference while producing identical outputs.
\end{abstract}

\maketitle

\section{Introduction}
%\shan{I have re-written the introduction all the way until the paragraph  `VeriCache handles both long-context decoding and remote prefix caching ...`. If you plan to make big changes to the paragraphs that I already edited, please discuss with me first.}

\begin{figure}[t]
    \centering
    \includegraphics[width=\columnwidth]{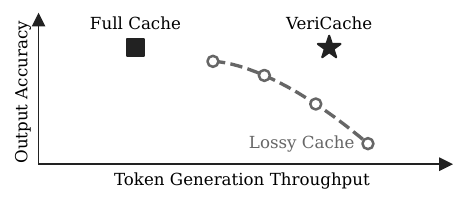}
    \caption{The accuracy--throughput dichotomy. \name attains the throughput of compression with the same output as full KV.}
    \label{fig:intro_tradeoff}
\end{figure}

The context lengths of state-of-the-art
LLMs have grown to more than one million tokens. This growth has powered many applications, from repository-level code generation~\cite{llm4code1, llm4code2, llm4code3, llm4code4, llm4code5}, multi-document reasoning~\cite{llm4multi1, llm4multi2, llm4multi3, llm4multi4}, to agentic workflows with long interaction histories~\cite{llm4agent1, llm4agent2, llm4agent3, llm4agent4}.

The performance impact of KV cache manifests at both single-request and multi-request LLM serving.
Within a single request, every decoding step must read the entire KV cache from GPU memory (HBM) to on-chip memory for attention.
Furthermore, the size of the KV cache degrades request throughput, as it reduces the number of requests that can fit in GPU memory~\cite{shadowkv, lmcache}.
Across requests, KV cache reuse is a common practice to avoid repetitive KV cache generation among requests that
share long common prefixes (e.g., system prompts, shared documents, prior conversation turns). Then large KV caches on storage need to be quickly loaded onto GPU memory when a request arrives, which can dominate request latency~\cite{cachegen, lmcache, shadowserve}.

A growing line of work tackles these problems by \emph{compressing} the KV cache---either by dropping tokens at selected layers and attention heads~\cite{duoattention, fastkvzip, kvzap, kvpress, quest} or by reducing precision through quantization~\cite{cachegen, kvquant, kivi, turboquant, llm265}.
Both deliver substantial efficiency gains---with 2--5$\times$ reductions in memory or transfer size.
%{I guess you also need to mention cartridges... Expert reviewers would wonder how trained compressed KV performs in terms of error propagation. } \jiayi{i think that's just one of the token dropping methods if we generalize the term a bit like token merging}

Although effective at improving throughput, compressing KV cache changes its content
%\footnote{Lossless compression is rarely used because it offers limited throughput gains.}
, causing inference output to diverge from the distribution of decoding outputs using the full-size KV cache.
As we will show in Section~\ref{sec:motivation}, 
the probability of divergence accumulates with more output tokens. 
When an application needs long output, the output divergence often violates
functional correctness or structural requirements, even if the output is still smooth natural language.
%the token-level accuracy metrics typically used to evaluate these methods---F1, ROUGE, perplexity---are forgiving of small per-token deviations, and 
%do not capture the functional correctness that structured-output tasks require.
On coding and tool-calling benchmarks, functional accuracy (e.g., syntax validity, exact argument matching) degrades sharply even at moderate compression ratios. 
%, despite token-level metrics remaining high.

%The root cause is that KV compression introduces a per-token distribution shift that accumulates over the generated sequence, making longer outputs increasingly likely to diverge from the full-KV output.
This creates a dichotomy (Figure~\ref{fig:intro_tradeoff}): accept lossy KV and risk output quality, or use full KV at the cost of much lower throughput. 
We pose the following question:

\begin{center}
\emph{Can we exploit the throughput benefits of KV cache compression {without affecting LLM output}?}
\end{center}

In this paper, we present \name, a new inference scheme inspired by speculative decoding~\cite{eagle1, eagle2, kdspectemperature, fastmtp, specngram}.
Instead of directly serving tokens from compressed KV cache, \name uses the compressed KV cache to \emph{draft} tokens, then \emph{verifies} them against the full KV cache.
Wrong tokens are corrected, so the final output is identical \footnote{In this paper, we define \textit{identical} results as identical under greedy decoding (i.e., zero temperature) except for randomness caused by hardware} to full-KV inference.

Directly applying existing speculative decoding techniques is insufficient as token verification overhead can cause throughput degradation (see \S\ref{sec:veri}), negating the benefits of KV cache compression.
\name uses two key designs that take advantage of the fact that drafting and verification in \name use \textit{exactly} the same model and model weights, a property different from traditional speculative decoding.

First, \name schedules \textit{cross-resource staggering}.
By sharing exactly the same model (weights), token drafting and token verification become feasible to execute in the same batch, offering a huge performance benefit as drafting and verification have \emph{complementary resource bottlenecks}.
Drafting decodes one token at a time using compressed KV located in GPU memory. 
Its sequential vector-matrix multiplication under-utilizes GPU FLOPs and makes it
a GPU memory bandwidth bound operation.
Verification, in contrast, requires loading the full KV cache from secondary storage into GPU and verifying multiple tokens in parallel, putting its bottleneck on
inter-connect bandwidth and GPU FLOPs.
Consequently, staggering drafting and verification across these distinct hardware resources offers much better utilization than running only drafting or only verification in each batch---the lock-step playbook in traditional speculative decoding. 
Cross-resource staggering requires much more sophisticated scheduling, which we will explain in \S\ref{sec:design}.

Second, \name uses an extended verification period that 
maximizes the parallelism of verification and minimizes the loading frequency of full KV cache. 
How often verification happens (i.e., every how many output tokens per verification) depends on how likely the drafted tokens would be accepted by the verification process. 
The more likely they are, the less frequent the verification can be. 
Different from traditional speculative decoding,
compressed KV caches retain the same model weights and dominant attention patterns as the full model, so the compressed-KV drafter sustains much longer accepted runs than a traditional small-model drafter---25--40 accepted tokens per verification round for \name vs.\ only 2--3 for typical small-model drafters (Section~\ref{subsec:p2}).

\name handles both long-context decoding and remote prefix caching, with a runtime scheduler that adapts verification frequency and batch composition to hardware and workload conditions.

The idea of driving a speculative drafter from a sparser or compressed KV cache is not new. MagicDec~\cite{magicdec} pairs a small-model drafter with a sparse KV cache; QuantSpec~\cite{quantspec} and SparseSpec~\cite{sparsespec} extend it to self-speculation with hierarchical quantization and dynamic sparse attention, respectively. 

However, there are two key differences.
First, prior systems all keep the full KV cache in GPU memory, capping compression's throughput gains.
In long-context decoding, they cannot realize compression's batch-size or HBM-bandwidth benefits; \name instead dedicates HBM to compressed KV and reloads the full KV from host DRAM only at verification (\S\ref{subsec:p1}).
In remote prefix caching---where the bottleneck is the slow storage-to-remote-GPU link, not HBM---no prior speculative-with-compressed-KV system applies; \name's remote drafter sees only the compressed KV over the slow link, while a local GPU on the fast link either caches or loads and applies the full KV for verification.
In long-context decoding---the setting where prior systems apply---\name also achieves higher throughput (\S\ref{sec:eval}).

Second, \name enables lossless inference at a high throughput for a \textit{range} of lossy KV cache compression techniques---quantization and token dropping---for the first time. 
\name exposes a uniform compressor interface: any token-dropping or quantization method that conforms can serve as the drafter, with no change to scheduling, verification, or transfer (\S\ref{sec:interface}). 
We instantiate this interface for seven existing methods spanning token dropping and quantization, whereas prior systems hard-wire a single compressor.

\name, built on top of vLLM~\cite{vllm} and LMCache~\cite{lmcache}, achieves up to $4\times$ higher throughput than full-KV inference on long-context decoding and up to $2\times$ on remote prefix caching across models and workloads, with identical outputs.

\section{Background}

\begin{table}[!t]
\centering
\caption{Representative KV cache compression methods.}
\label{tab:approx_methods}
\small
\setlength{\tabcolsep}{6pt}
\renewcommand{\arraystretch}{1.18}

\begin{tabularx}{\columnwidth}{
  >{\raggedright\arraybackslash}p{0.27\columnwidth}
  >{\raggedright\arraybackslash}X
}
\toprule
\textbf{Strategy} & \textbf{Methods} \\
\midrule

\textbf{Token dropping}
&
Keyformer~\cite{keyformer},
H2o~\cite{h2o},
Ada-KV~\cite{adakv},
LagKV~\cite{lagkv},
KVzip~\cite{kvzip},
FastKVzip~\cite{fastkvzip},
KVzap~\cite{kvzap},
SnapKV~\cite{snapkv},
PyramidKV~\cite{pyramidkv},
PyramidInfer~\cite{pyramidinfer},
DuoAttention~\cite{duoattention}.
\\

\midrule

\textbf{KV quantization}
&
KIVI~\cite{kivi},
KVQuant~\cite{kvquant},
KVTuner~\cite{kvtuner},
TurboQuant~\cite{turboquant},
CacheGen~\cite{cachegen},
KVTC~\cite{kvtc},
QServe~\cite{qserve},
GEAR~\cite{gear},
LLM.265~\cite{llm265},
RotateKV~\cite{rotatekv}.
\\

\bottomrule
\end{tabularx}
\end{table}

\subsection{The KV cache bottleneck in inference}
LLM inference consists of two phases: (1) prefill, which generates the KV cache for a prompt, and (2) decode, which generates tokens autoregressively from the KV cache.
The KV cache introduces two overheads scaling with context length.

\textbf{Within-request KV overhead.}
The KV cache incurs $O(n)$ memory footprint and bandwidth cost that grows linearly with context length $n$~\cite{kvsurveyfromdell, kvsurvey1, kvsurvey2}.
On the memory side, larger KV caches leave less room for batching requests: serving Qwen-32B~\cite{qwen3report} (${\sim}$64GB weights) on a single H100 80GB GPU, a 2K-token context requires ${\sim}$0.3GB of KV per request, allowing a batch of ${\sim}$50 requests; scaling to 100K tokens grows the KV to ${\sim}$15GB, reducing the batch size to 1.
On the bandwidth side, each decode step must read the full KV from HBM: serving Llama-3.1-8B-1M~\cite{llamaultralong} (${\sim}$16GB weights) on an H100 (3TB/s HBM), each decode step takes ${\sim}$5ms at 5K context (${\sim}$0.6GB KV) and ${\sim}$25ms at 500K context (${\sim}$60GB KV)---decoding 100 tokens takes ${\sim}$0.5s at 5K context but ${\sim}$2.5s at 500K.

\textbf{Cross-request KV overhead.}
Many long-context workloads share long prefixes across requests---system prompts, shared documents, prior conversation turns---making the $O(n^2)$ prefill redundantly expensive when each request recomputes them from scratch.
To avoid this, systems precompute KV caches once and reuse them across requests~\cite{lmcache, sglang, kvflow, mooncake}.
However, reusing a precomputed cache requires transferring KV caches from storage or across the network onto the serving GPU, and this transfer can itself become the new bottleneck at long contexts~\cite{impress, cachegen, shadowserve, mooncake}.
For example, loading Qwen-32B's precomputed KV cache from S3 (${\sim}$3GB/s~\cite{s3performance}) takes ${\sim}$0.5s at 10K context (${\sim}$1.5GB) but ${\sim}$5s at 100K context (${\sim}$15GB)---matching or exceeding the prefill time reuse was meant to avoid.

\subsection{KV cache compression techniques}

A growing line of work tackles these bottlenecks by compressing the KV cache (Table~\ref{tab:approx_methods}).
These methods fall into two categories.
\textbf{Token dropping} changes the shape of the cache by dropping tokens at certain layers and attention heads.
For example, StreamingLLM~\cite{streamingllm} retains only a few initial ``attention sink'' tokens plus a sliding window of recent tokens, enabling unbounded generation.
DuoAttention~\cite{duoattention} identifies attention heads that need full context and drops most entries for the rest.
KVzip~\cite{kvzip} scores KV pair importance via a context reconstruction loss at prefill and evicts low-importance pairs once for reuse across queries.
FastKVzip~\cite{fastkvzip} learns a per-token gating function that evicts low-importance entries.
\textbf{KV quantization} preserves the cache shape but reduces its per-element precision---for example, KVQuant~\cite{kvquant} quantizes keys per-channel before rotary positional embedding to preserve outlier structure at sub-4-bit precision.
KIVI~\cite{kivi} applies per-channel quantization to keys and per-token to values for tuning-free 2-bit compression.
TurboQuant~\cite{turboquant} randomly rotates key/value vectors to induce a known coordinate distribution, enabling optimal per-coordinate scalar quantization at 2--4 bits without calibration.
CacheGen~\cite{cachegen} encodes KV tensors into compact bitstreams by exploiting token-wise locality and layer-wise sensitivity.

\textbf{Inherently lossy.} All these compression methods are inherently \emph{lossy}: the compressed KV cache deviates from the original, and these deviations propagate through inference.

% !TEX root = sample-sigconf.tex
\begin{figure}[!t]
    \centering
    \includegraphics[width=\columnwidth]{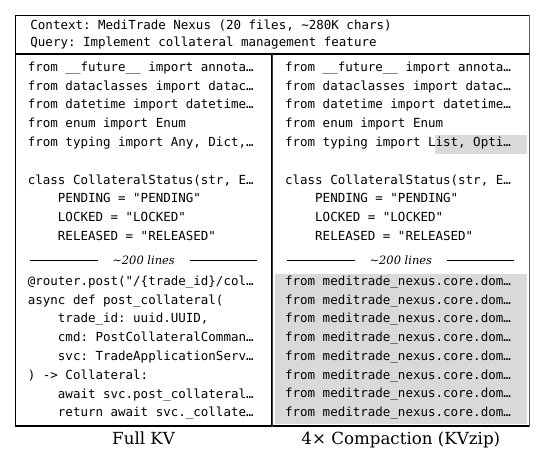}
    \caption{Code-generation failure from compressed KV. Asked to implement a feature over a $\sim$280K-character codebase, Full~KV produces correct code while KVzip 4$\times$ goes clearly wrong after $\sim$200 lines.}
    \label{fig:text_comparison}
\end{figure}

\section{Motivation: Why Lossy KV Methods Fail}\label{sec:motivation}

\subsection{Semantic similarity $\neq$ functional correctness}

\begin{figure}[!t]
    \centering
    \includegraphics[width=\columnwidth]{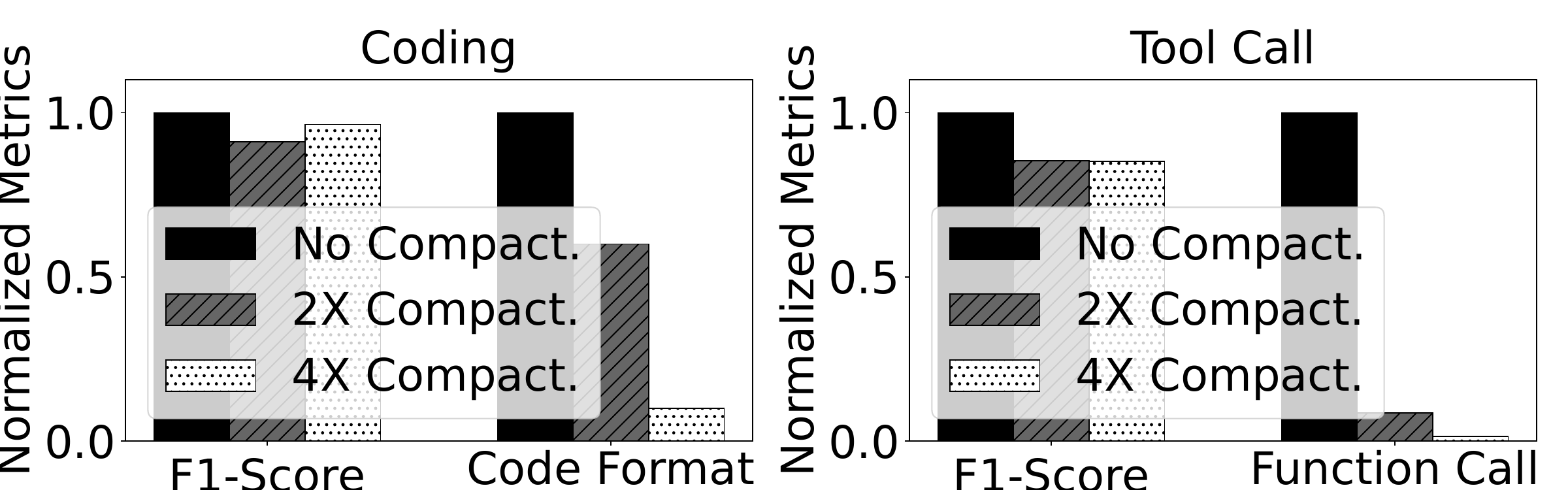}
    \caption{Soft metrics mask functional failures (all metrics normalized to no-compression baseline).}
    \label{fig:motivation_metrics}
\end{figure}

Previous KV compression techniques have been evaluated using 
token-level metrics~\cite{keyformer,h2o,duoattention,kivi,lserve,turboquant}, such as F1, ROUGE, perplexity, and cosine similarity. 
These token-level metrics tolerate small text deviation and are suitable for open-ended or short-answer tasks such as summarization, natural language Q\&A, etc. 
However, they are ill-fitted for applications that have strict syntax or semantics requirement.
For example, an invalid syntax token can break a program; a misplaced delimiter can corrupt a JSON object; and a wrong operator alters the semantics of a shell command (e.g., \texttt{rm -rf *.logs} $\to$ \texttt{rm -rf * .}). 
Figure \ref{fig:text_comparison} shows a real code-generation failure caused by the use of compressed KV.

Figure~\ref{fig:motivation_metrics} illustrates this significant difference between a token-level metric (F1) and functional metrics (i.e., code format accuracy and function call accuracy).
These two types of metrics are measured when we apply a representative token-dropping method, KVzip~\cite{kvzip}, on Qwen3-Coder-30B \cite{qwen3report}.
The left figure shows the results on a coding task in SWE-bench Lite~\cite{swebench}; the right figure shows the results on a tool-calling task in ComplexFuncBench~\cite{complexfuncbench}.

The pattern is consistent across both tasks: F1, which gives partial credit for partially correct outputs, remains above 75\%, but functional correctness collapses~\cite{yourChatGPTCorrect}.
Code format accuracy---which requires the output to be a syntactically valid git diff---drops to near zero under KVzip 4$\times$ compaction.
Function call accuracy---which requires every call name and argument to match exactly---drops below 10\% under KVzip 4$\times$ compaction.

For some of the fastest-growing LLM use cases including code generation, agentic tool use, and any tasks that require structured output, traditional token-level metrics are too lenient: any partial token-level deviation could be devastating. 

\subsection{Root cause: per-token bias accumulation}
The quality collapse shown in Figure \ref{fig:motivation_metrics} reflects a fundamental limitation of KV compression.
The altered KV cache changes the attention weights at every layer. It replaces the model's learned next-token distribution $p_\text{full}$ with a different distribution $p_\text{lossy}$ that the model was never trained to produce.
Unlike sampling noise (e.g., temperature), which draws different tokens \emph{within} $p_\text{full}$, this is a systematic bias: every sample is drawn from the wrong distribution, and no amount of resampling corrects it.

\paragraph{Per-step distribution shift.}
We quantify this bias with the KL divergence~\cite{introtoinfotheory} between the full-KV and lossy-KV next-token distributions at each decoding step $t$:
\begin{equation}
    \mathrm{KL}_t \;=\; \sum_{x_t} p_\text{full}(x_t\mid x_{<t}) \,\log \frac{p_\text{full}(x_t\mid x_{<t})}{p_\text{lossy}(x_t\mid x_{<t})}.
    \label{eq:kl_step}
\end{equation}
$\mathrm{KL}_t$ is zero only when the two conditional distributions are identical, and grows as they diverge.

\begin{figure}[t]
    \centering
    \includegraphics[width=\columnwidth]{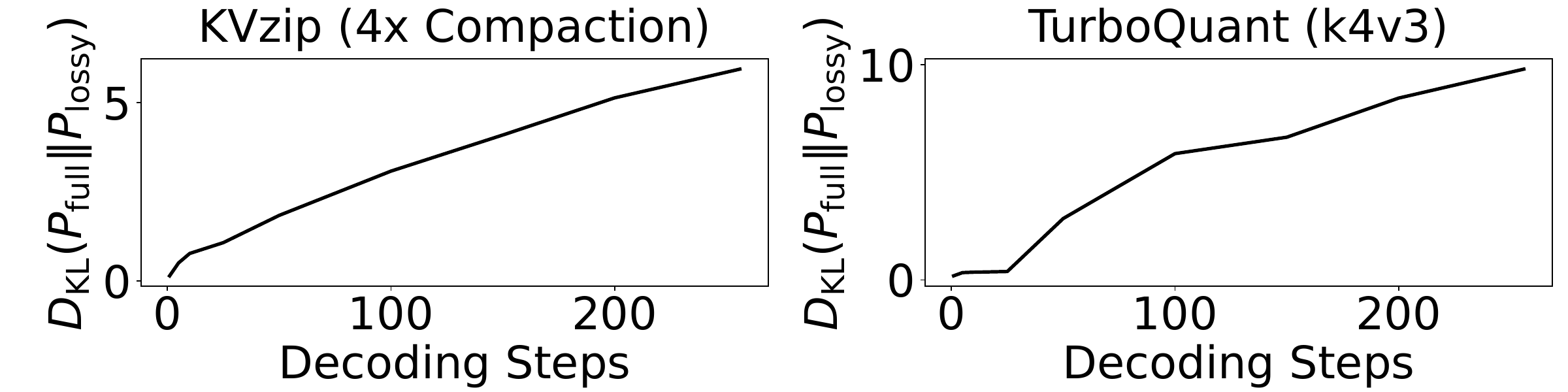}
    \caption{Sequence-level KL $\mathrm{KL}(p_\text{full}(x_{1:t})\,\|\,p_\text{lossy}(x_{1:t}))$ grows roughly linearly in $t$ under KVzip 4$\times$ (left) and TurboQuant k4v3 (right); at temperature 0.5.}
    \label{fig:motivation_kl}
\end{figure}

\paragraph{Distribution shift compounds over the generation.}
This per-step bias accumulates over the autoregressive generation.
By the chain rule of KL divergence:
\begin{equation}
    \mathrm{KL}_{1:T} \;\triangleq\; \mathrm{KL}\!\left(p_\text{full}(x_{1:T}) \,\|\, p_\text{lossy}(x_{1:T})\right)
    \;=\; \sum_{t=1}^{T} \mathbb{E}_{x_{<t}\sim p_\text{full}}\!\left[\mathrm{KL}_t\right].
    \label{eq:kl_chain}
\end{equation}
if per-step KL exceeds $\varepsilon > 0$, sequence-level KL grows linearly: $\mathrm{KL}_{1:T} \geq \varepsilon T$. Since $\mathrm{KL}_{1:T}$ equals
$\mathbb{E}_{x \sim p_\text{full}}\!\big[\log(p_\text{full}(x_{1:T})/p_\text{lossy}(x_{1:T}))\big]$, this means that for a sequence sampled from $p_\text{full}$, the log-likelihood ratio has mean ${\geq}\,\varepsilon T$, and so the likelihood ratio $p_\text{full}(x_{1:T})/p_\text{lossy}(x_{1:T})$ is typically of order $e^{\varepsilon T}$---i.e., \emph{exponential} in $T$ (proof in Appendix~\ref{app:kl_chain}). Figure~\ref{fig:motivation_kl} confirms this empirically: $\mathrm{KL}_{1:T}$ grows linearly with decoding steps under both KVzip 4$\times$ compaction and TurboQuant k4v3 quantization.

Concretely, in Figure~\ref{fig:motivation_kl}, KVzip 4$\times$ accumulates only ${\sim}0.023$ nats of KL per step---so the lossy model assigns the full-KV token $e^{-0.023}{\approx}98\%$ of its full-KV probability, barely distinguishable.
After $T{=}250$ steps, cumulative KL reaches ${\sim}6$ nats, so the lossy model emits the full-KV output sequence with probability only $e^{-6}{\approx}2.5{\times}10^{-3}$---the ${\sim}2\%$ per-step gap amplified into a $400\times$ mismatch over a few hundred tokens.
Figure~\ref{fig:text_comparison} shows the consequence: on a long code generation task, the 4$\times$ compacted output starts correctly but rapidly degenerates as compounding bias drives the model off its learned distribution.
%\shan{is this coding example real? can you show a real example?} \jiayi{this is real}

%This exponential blow-up of the likelihood ratio is the root cause of the functional accuracy collapse in Figure~\ref{fig:motivation_metrics}: structured outputs require many tokens to be jointly correct, and each additional token multiplies the gap between the lossy and full distributions.

\paragraph{The accuracy-efficiency dichotomy.}
For precision-critical applications, existing KV compressions force a binary choice: accept lossy KV and risk silent failures, or use full KV and forgo the efficiency gains.
We argue that this dichotomy is unnecessary: compression should not replace exact computation, but \emph{accelerate} it without sacrificing correctness.

%\kt{An overview section here}

\section{KV Cache Verification}
\label{sec:veri}
\subsection{\name overview}\label{subsec:overview}

\begin{figure}[t]
    \centering
    \includegraphics[width=\columnwidth]{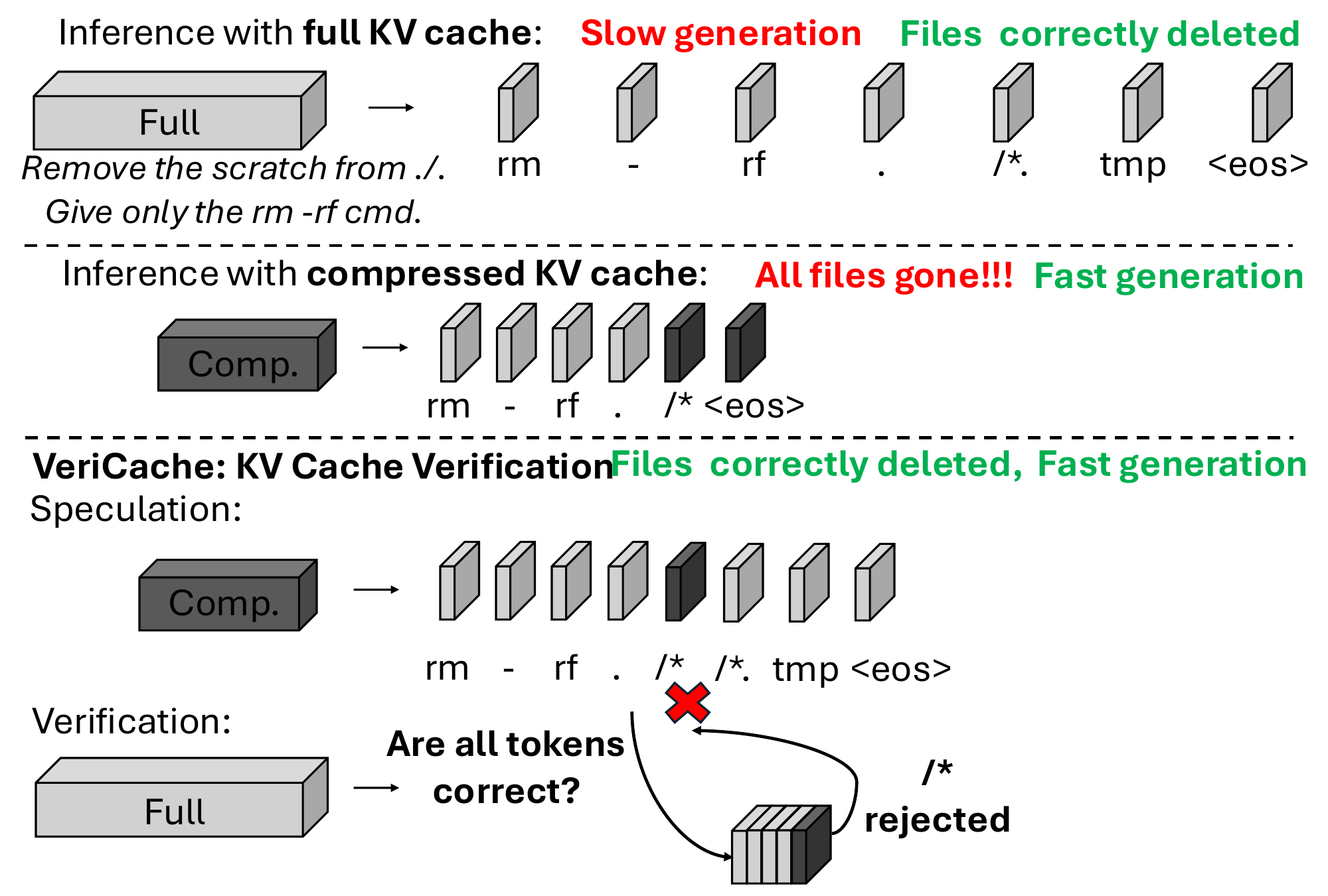}
    \caption{Overview of \name. Tokens drafted with compressed KV, then verified with full KV.}
    \label{fig:baseline_illustrate}
\end{figure}

At the algorithm level, \name repurposes any lossy KV compression method into a speculative execution layer of the inference pipeline, where tokens are drafted quickly. 
These drafted tokens are then verified against the full KV cache to guarantee correctness (Figure~\ref{fig:baseline_illustrate}).
Concretely, let $\text{KV}_\text{comp}$ denote the compressed KV cache and $\text{KV}_\text{full}$ the full cache. 
The core loop proceeds as follows\footnote{We describe greedy decoding for clarity; \name extends to sampling-based decoding via standard rejection sampling~\cite{eagle1, kdspectemperature}.}:

(1)~\textit{Draft}: autoregressively generate $x$ candidate tokens $t_1,\dots,t_x$ using $\text{KV}_\text{comp}$---each $t_i$ is the most-likely next token under the model's distribution conditioned on the prompt (represented by $\text{KV}_\text{comp}$) and the previously generated tokens.

(2)~\textit{Verify}: one forward pass over the $x$ drafted positions in parallel, conditioned on $\text{KV}_\text{full}$ and $t_{1..k-1}$ at each position $k$. A single forward pass yields $x{+}1$ predictions $t_1^*,\dots,t_{x+1}^*$: one at each drafted position $k$ (the full-KV next-token prediction given $t_{1..k-1}$), plus one bonus prediction $t_{x+1}^*$ that follows the last drafted token.

(3)~\textit{Accept}: find the first position $j \in [1, x]$ where $t_j \neq t_j^*$; accept $t_1,\dots,t_{j-1}$ and the verifier's correction $t_j^*$, and discard the rest. If all $x$ drafted tokens match (no such $j$ exists), accept all $x$ of them plus the bonus $t_{x+1}^*$. Drafting then resumes from the position immediately after the last accepted token.
%---so each verify round produces $1$ to $x{+}1$ accepted tokens.

%Because verification uses the full KV cache, \name produces output identical to standard decoding. %\jiayi{junchen: say we use the same model}

At the system level, each verify step demands three resources beyond drafting: (1) \emph{interconnect bandwidth} to load $\text{KV}_\text{full}$ from CPU or remote storage, (2) \emph{GPU HBM} to hold it during the verify pass (competing with resident compressed caches), and (3) \emph{GPU compute} for the verify pass over $x$ tokens. The first two are tightly budgeted and lock-step verification would spike both; compute is handled implicitly, since staggering verifies across iterations (\S\ref{sec:design}) smooths the compute. Without care, these costs would erase compression's throughput gain. We next present two design principles (P1, P2) that let \name pay for verification without giving it back.
%Specifically, \name follows two design principles:
%\begin{itemize}[leftmargin=*,itemsep=2pt]
%    \item \textbf{P1 (cross-resource staggering).} In \name, drafting and verification both carry out forward passes using exactly the same model, and yet consume different hardware resources. Therefore, \name staggers each request's verify across iterations: at every iteration some requests are drafting while others are verifying, keeping all resources busy \shan{What is a `request`? a bit confused}(\S\ref{subsec:p1}).
%    \item \textbf{P2 (high acceptance rate, long drafts).} \name's drafter is the target model itself running on compressed KV, so drafted sequences track the full-KV output closely; \name sustains acceptance lengths of $25$--$40$ tokens per round, so verification fires rarely and its amortized cost is low (\S\ref{subsec:p2}). \jiayi{add adaptation if time}
%\end{itemize}
The runtime mechanisms that realize these principles are detailed in Section~\ref{sec:design}.

\subsection{P1: Cross-resource staggering}\label{subsec:p1} % alias for back-compat with intro / design refs

\begin{figure}[t]
    \centering
    \includegraphics[width=\columnwidth]{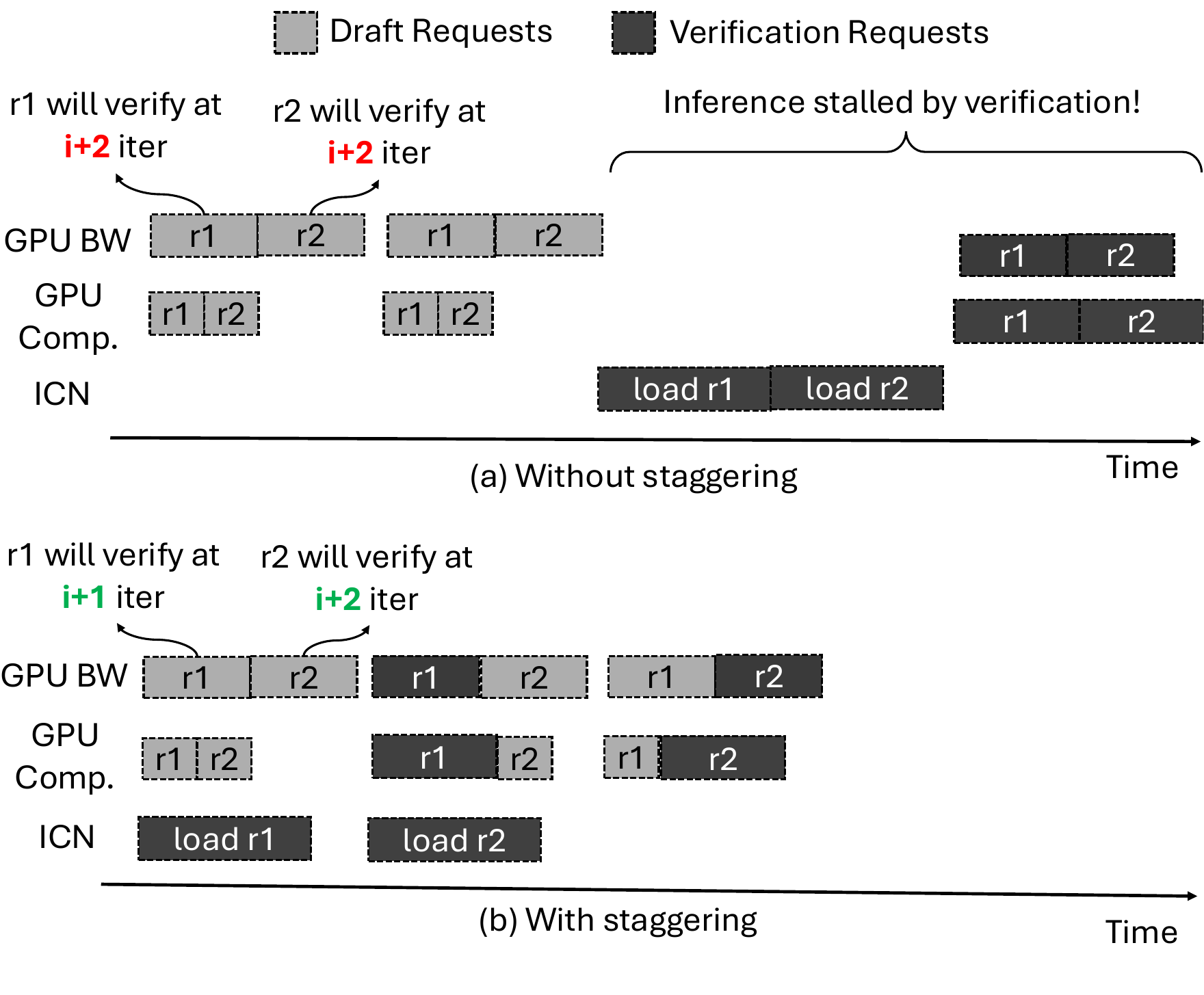}
    \caption{Cross-resource staggering. \textbf{(a) Lock-step:} simultaneous verification at $i{+}2$ congests the interconnect and stalls the GPU. \textbf{(b) Staggered:} shifting $r_1$ to $i{+}1$ overlaps KV transfer and verify with the other request's draft, keeping interconnect, GPU, and HBM busy.}
    \label{fig:phase_mix_illustrate}
\end{figure}

Traditional speculative decoding runs requests in lock-step: all draft for $x$ iterations, then all verify at iteration $x{+}1$. Every iteration shares the same bottleneck, leaving other resources idle.

\name's first design principle is to stagger requests instead, mixing some requests' verification into iterations where others are drafting. Since compressed-KV drafting and full-KV verification consume complementary resources (elaborated below), staggering yields much better resource utilization. We describe the design for both use cases below; the runtime mechanics appear in \S\ref{sec:design}.

\subsubsection{Staggering with long-context decoding}

\paragraph{Deployment and resource consumption} 
Traditionally, when compressed KV is used for long-context decoding, the KV cache is kept in compressed form in GPU high-performance memory (HBM) throughout serving, and the full KV is never retained on the GPU.
In subsequent iterations, the compressed KV cache is read from HBM to help generate the next token(s).

\begin{figure}[t]
    \centering
    \includegraphics[width=\columnwidth]{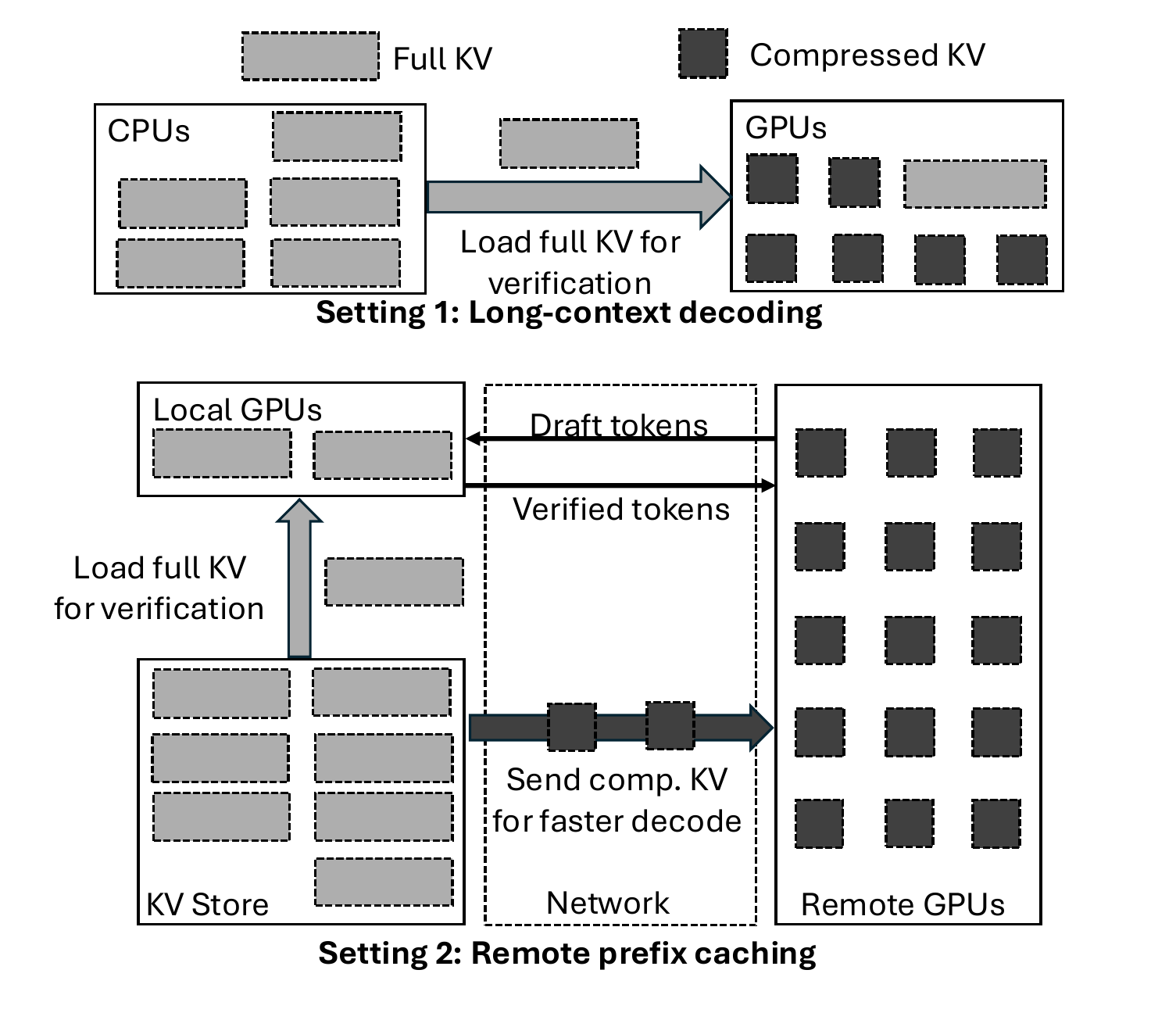}
    \caption{\name's two settings: long-context decoding (top, full KV on CPU) and remote prefix caching (bottom, full KV in local storage).}
    \label{fig:setting}
\end{figure}

In \name, we keep all the traditional setting and in addition, keep the full in CPU memory as shown in Figure \ref{fig:setting}.
Whenever this request's drafted tokens are ready to be verified, the full KV is reloaded from CPU to GPU over the CPU--GPU interconnect, and used for verification.

%\paragraph{Observation 1: Drafting and verification hit distinct hardware resources.}
Under this setting, the performance of drafting is bounded by HBM-bandwidth: the GPU spends an iteration to read model weights and the compressed KV cache from HBM, with little compute (i.e., forward pass for one token) to do.
In contrast, the performance of verification draws on different resources entirely: it must transfer $\text{KV}_\text{full}$ across an interconnect (CPU--GPU) and run a forward pass over $x$ drafted tokens, putting its bottleneck on interconnect bandwidth and GPU compute rather than HBM bandwidth.

%at the verify step, all $B$ $\text{KV}_\text{full}$ transfers congest the interconnect and all $B$ verification forward passes fire at once on GPU compute, while both resources sit idle during the drafts (Figure~\ref{fig:phase_mix_illustrate}a). The distinct resources are never busy at the same time.

\paragraph{Staggering schedule \& performance estimation}
We now discuss our staggering schedule in more details, together with a performance estimation; the full derivations with relaxed assumptions are in Appendix~\ref{app:theoretical_analysis}. Table~\ref{tab:notation} summarizes the notation.

\begin{table}[t]
\centering
\caption{Notation for throughput analysis.}
\label{tab:notation}
\footnotesize
\setlength{\tabcolsep}{4pt}
\begin{tabular}{l l}
\toprule
\textbf{Symbol} & \textbf{Description} \\
\midrule
$M$ & Model weights size \\
$\text{KV}_\text{full}$ & Full KV cache size per request \\
$c$ & Compression ratio ($\text{KV}_\text{comp} = c \cdot \text{KV}_\text{full}$, $c < 1$) \\
$\gamma(x,c)$ & Acceptance rate (fn.\ of draft length $x$ and compression $c$) \\
$x$ & Draft (speculation) length \\
$B$ & Batch size (number of concurrent requests) \\
$K$ & Number of output tokens per request \\
$\text{BW}_\text{hbm}$ & GPU HBM bandwidth \\
$\text{BW}_\text{inter}$ & CPU--GPU interconnect bandwidth \\
$\text{BW}_h$, $\text{BW}_l$ & Storage--local / storage--remote bandwidth \\
\bottomrule
\end{tabular}
\end{table}

Consider a single GPU serving $B$ long-context requests. 
Each request keeps $\text{KV}_\text{comp} = c \cdot \text{KV}_\text{full}$ on GPU and $\text{KV}_\text{full}$ on CPU. 
Cross-resource staggering schedules verifications so that, at any iteration, only about $B/x$ of the $B$ requests are in the verify phase, each checking the $x$ tokens it most recently drafted. 
Per iteration, the GPU does $B$ single-token forward passes for the drafting requests plus ${\sim}(B/x) \cdot x = B$ token forward passes for the staggered verifications---roughly $2B$ tokens total. 
The HBM traffic these passes incur has three components: the model weights ($M$), the compressed KV cache of every drafting request ($B \cdot c \cdot \text{KV}_\text{full}$, the compressed-KV bandwidth used by drafting), and the full KV cache of the $B/x$ verifying requests ($(B/x)\cdot \text{KV}_\text{full}$, the full-KV bandwidth used by verification). 
In parallel, the next $B/x$ full KV caches transfer from CPU to GPU over the interconnect, overlapping with the forward passes. 
The iteration time $T_\text{iter}$ is therefore the longer of the GPU forward-pass time $T_\text{gpu}$ and the CPU--GPU transfer time $T_\text{xfer}$:

\begin{equation}\label{eq:t_iter}
    T_\text{iter} = \max\!\Big(\underbrace{\frac{M + B \cdot \text{KV}_\text{full} \cdot (c + 1/x)}{\text{BW}_\text{hbm}}}_{T_\text{gpu}},\;\; \underbrace{\frac{B \cdot \text{KV}_\text{full}}{x \cdot \text{BW}_\text{inter}}}_{T_\text{xfer}}\Big)
\end{equation}

This staggered schedule offers better resource utilization than the traditional lock-step schedule that conducts all-draft iterations followed by all-verify iterations: the all-draft iterations leave the CPU-GPU link idle; even if CPU-GPU transfer overlaps with drafting iterations, the resource utilization of all-verify iterations would still be a concern. Assume that requests $B_1$, $B_2$ ... $B_k$ are scheduled to enter the same all-verify iteration. 
Transferring their full KVs from CPU to GPU will take a lot of time, much longer than one forward-pass iteration. 
Even worse, since the transfer is serial, the full KV of $B_1$ will have to stay in the HBM idle, waiting for the full KVs of $B_2$, ... $B_k$ to arrive before the verification starts. 
We further explain this comparison with a concrete example below.

\emph{Concrete example.} Consider serving Mistral 24B on an RTX PRO 6000 (PCIe Gen5 $\times16$, 64 GB/s) with $B{=}10$ requests, $\text{KV}_\text{comp}{=}1$ GB and $\text{KV}_\text{full}{=}4$ GB per request, and draft length $x{=}30$. 
Each verify transfers one $\text{KV}_\text{full}$ over PCIe ($\sim80$ ms).
A draft-only iteration reads $M + B\cdot\text{KV}_\text{comp}$ from HBM ($\sim35$ ms); folding in a verify adds one $\text{KV}_\text{full}$ to the HBM read, extending it to $\sim37$ ms.
With \name, the $10$ verifies are spread one every $3$ draft iterations, so each $\sim80$-ms PCIe transfer overlaps with concurrent draft work and peak HBM stays at $M + B\cdot\text{KV}_\text{comp} + 1\cdot\text{KV}_\text{full} = 64$ GB. 
A lock-step alternative that batches all $10$ verifies at iteration $30$ serializes $40$ GB on the single PCIe link ($\sim800$ ms of transfer time, $\sim20\times$ the iteration window) and spikes HBM to $M + B\cdot\text{KV}_\text{full} = 90$ GB (assuming compressed cache is offloaded to CPU). 
A sequential variant (one verify per iteration after drafting) avoids the HBM spike but still pays the same $\sim800$ ms of PCIe transfer overhead per $30$-token draft cycle that staggering eliminates.

\subsubsection{Staggering with remote prefix caching}
\paragraph{Deployment and resource consumption}
Traditionally, in remote prefix caching, pre-computed KV caches are saved on a storage node. When an incoming request's prefix matches one, the KV is transferred to a remote serving GPU before decoding can begin. Because the serving GPUs connect to the storage node over a slow link, transferring the compressed KV greatly reduces this transfer time.

\name additionally leverages the storage node's local GPU for verification (Figure~\ref{fig:setting}): on a prefix match, the storage node sends the compressed KV through the slow link to the remote GPU for drafting, and the full KV through the fast link to its local GPU for verification. Drafting and verification thus run on completely different hardware.

\paragraph{Staggering schedule \& performance estimation}
 %Consider a cluster where local GPUs have fast access to the storage node (bandwidth $\text{BW}_h$) while remote GPUs are connected via a slower link (bandwidth $\text{BW}_l \ll \text{BW}_h$). Without \name, a remote GPU must download $\text{KV}_\text{full}$ over the slow link before decoding, so latency per request is $\text{KV}_\text{full}/\text{BW}_l + K \cdot T_\text{decode}$ for $K$ output tokens. With \name, the remote GPU downloads only $\text{KV}_\text{comp}$ (${\sim}1/c\times$ faster) and starts drafting immediately, while a local GPU with fast storage access loads $\text{KV}_\text{full}$ and iteratively verifies batches of $x$ draft tokens.
Drafting (remote, HBM-bound) and verifying (local, storage-bandwidth-bound load plus a forward pass) use distinct hardware, so \name pipelines them within a single request as follows. 
Every $x$ drafted tokens trigger one verify: the remote drafter advances $x$ tokens on the compressed KV cache while, \emph{in parallel}, the local verifier prefetches the full KV cache from storage over $\text{BW}_h$. 
When both finish, the local runs a forward pass over those $x$ positions; the remote then waits for the accept/reject result before drafting the next $x$ tokens. 
So per-request, only the verify \emph{transfer} hides behind drafting---the verify forward pass itself sits on the critical path, because the next $x$ drafts depend on which of the previous $x$ were accepted. 
At the system level, while the remote pool waits on request~$r$'s verify forward pass, it drafts for other requests in the batch, so no GPU is ever idle. 
The per-request time is:
\begin{equation}\label{eq:tp_inter}
    T_\text{req} = \underbrace{\frac{c \cdot \text{KV}_\text{full}}{\text{BW}_l}}_{\text{startup}} + \underbrace{\frac{K}{x\,\gamma(x,c)}}_{\text{\# draft-verify cycles}} \cdot T_\text{cycle}
\end{equation}
where $T_\text{cycle} = \max\!\big(x \cdot T_\text{decode},\; \text{KV}_\text{full}/\text{BW}_h\big) + T_\text{fwd}(x)$: the $\max$ captures the draft--load overlap, and $T_\text{fwd}(x)$ is paid serially after it. The startup is ${\sim}1/c\times$ faster than the full-KV baseline ($\text{KV}_\text{full}/\text{BW}_l$), and $\gamma(x,c)$ directly reduces the number of draft-verify cycles needed.

\subsection{P2: High acceptance rate amortizes verification}\label{subsec:p2}

\paragraph{Observation 2: $x$ and $\gamma$ together govern verification cost.}
Given a compression ratio $c$, the draft length $x$ sets how often verification fires (once per $x{+}1$ iterations) and the acceptance rate $\gamma(x,c)$ sets how many drafted tokens each round actually keeps. 
If $x$ is too small, verification fires too often and the verification overhead dominates; if $\gamma$ is too low, most drafted tokens are discarded and the work in each round is wasted. 
Verification amortizes cheaply only when both $x$ and $\gamma$ are high.

\begin{insight}
\emph{Because \name's drafter is the target model itself running on the compressed KV cache, the compressed cache preserves the model's weights and the dominant attention patterns. 
The drafted tokens therefore closely track the full-KV output, so $\gamma$ stays high even at long draft lengths---verification fires rarely and a large fraction of each round's drafted tokens are accepted.}
\end{insight}

\begin{figure}[t]
    \centering
    \includegraphics[width=\columnwidth]{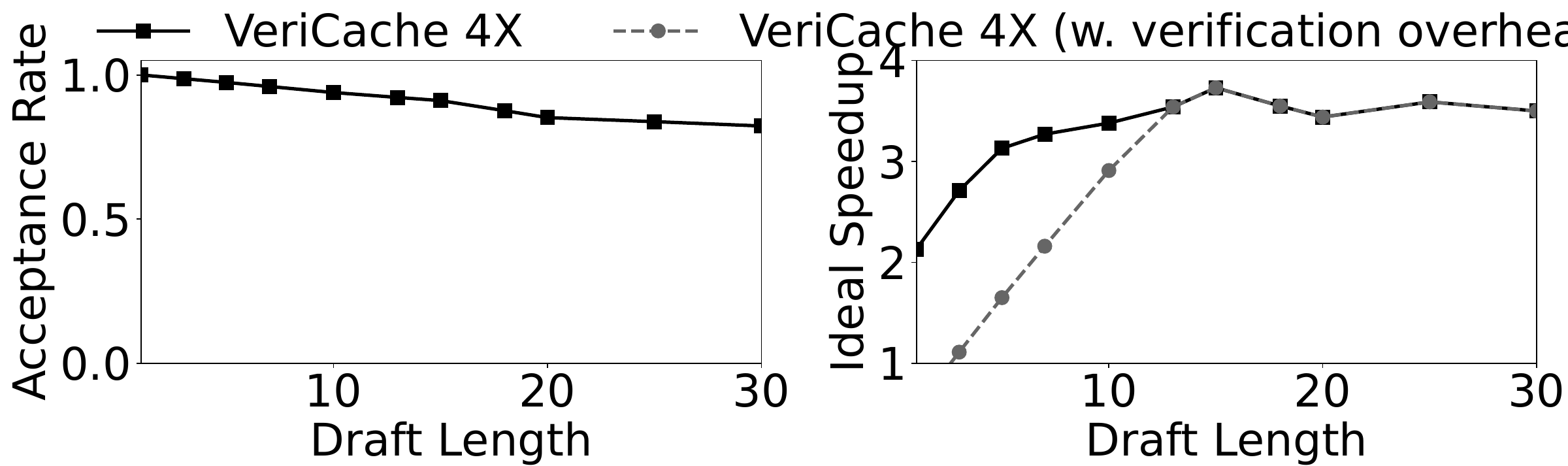}
    \caption{Acceptance rate (left) and ideal speedup (right) vs.\ draft length for \name at $4\times$ compaction.}
    \label{fig:key_insight_2}
\end{figure}

Figure~\ref{fig:key_insight_2} confirms this on 20 samples from the LMCache agentic trace~\cite{lmcache_trace}, running Llama-70B on 2$\times$H100 NVL under KVzip~\cite{kvzip} compaction. The left panel shows that at $4\times$ compaction, the acceptance rate stays above ${\sim}0.8$ even at draft length $30$---each verification round keeps a large fraction of its drafted tokens.
The right panel translates this into ideal speedup\footnote{Computed using the detailed throughput model in Appendix~\ref{app:theoretical_analysis}, which accounts for pipelined reloads and GPU memory constraints; Eq.~(\ref{eq:opt_app}) is a simplified version capturing the same tradeoffs.}: \name reaches a peak of ${\sim}3.7\times$ at draft length $15$. 
The dashed curve includes the verification overhead (after request staggering) and lies slightly below the solid (no-overhead) upper bound, with the residual gap---visible mostly at short draft lengths.

% \paragraph{Observation 3: The $\gamma(x,c)$ curve varies across requests.}
% At a fixed compression ratio $c$, the acceptance rate as a function of draft length is not a single curve: different requests have measurably different $\gamma(x,c)$ curves, so a globally chosen $x$ leaves performance on the table for many of them. \fillme \jiayi{add evidence: per-request $\gamma$ curve measurement}
%
% \begin{insight}
% \emph{A request's $\gamma$ behavior can be predicted from its first few drafted-and-verified tokens, so \name adapts the draft length $x$ \textbf{per request} from that online signal. Requests with high early acceptance get a longer $x$; requests with low early acceptance get a shorter $x$ or are demoted to non-speculative full-KV decoding. Section~\ref{sec:design} details the estimator and policy.}
% \end{insight}
% \jiayi{need to verify this quick and add some empirical numbers}

\begin{figure}[t]
    \centering
    \includegraphics[width=\columnwidth]{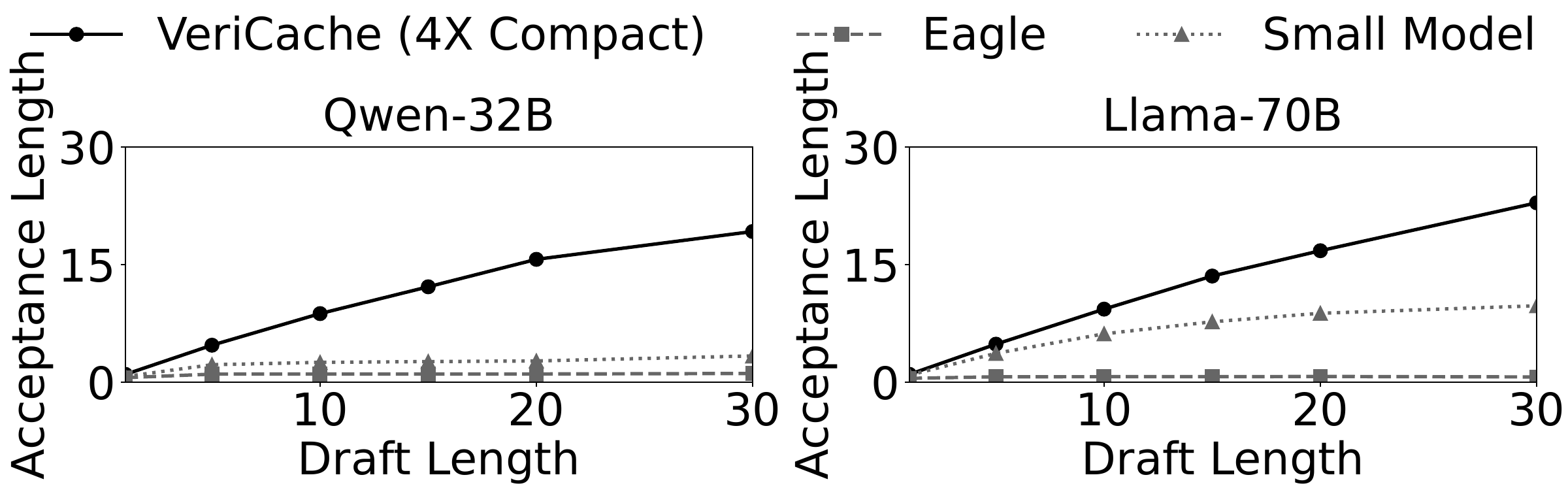}
    \caption{Acceptance length vs.\ draft length, comparing \name with Eagle and small draft model (Qwen-1.7B for Qwen-32B; Llama-8B for Llama-70B).}
    \label{fig:benefit1}
\end{figure}

By comparison, traditional speculative decoding drafts tokens from a small auxiliary model whose parameters differ from the target; because the two distributions diverge quickly, typical small-model drafters sustain only a few accepted tokens~\cite{speclen1, speclen2, specoverview}. 
\name keeps the target model and merely swaps in compressed KV, which is why its acceptance length stretches an order of magnitude further. Figure~\ref{fig:benefit1} quantifies the gap on Qwen-32B and Llama-70B: at draft length $30$, \name at $4\times$ compaction sustains acceptance lengths of ${\sim}19$ on Qwen-32B and ${\sim}23$ on Llama-70B, while Eagle saturates near $1$--$2$ on both, and a small draft model reaches only ${\sim}3$ on Qwen-32B and ${\sim}10$ on Llama-70B.

\begin{figure}[t]
    \centering
    \includegraphics[width=\columnwidth]{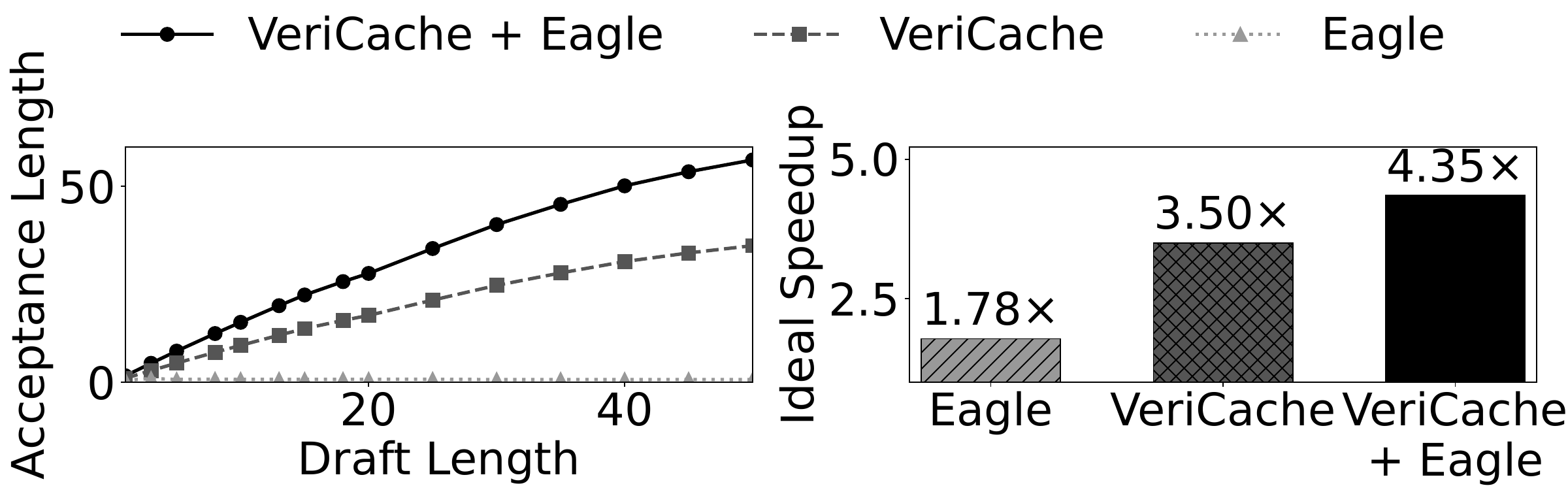}
    \caption{Composing \name with Eagle. (Left) Acceptance length vs.\ draft length. (Right) Ideal speedup.}
    \label{fig:benefit3}
\end{figure}

The two drafting strategies compose naturally: a small-model drafter (e.g., MTP~\cite{fastmtp}, EAGLE~\cite{eagle1}) proposes tokens that \name verifies against the compressed KV, periodically rechecking against the full KV to correct compression errors.
Figure~\ref{fig:benefit3} shows the payoff: \name~+~Eagle reaches $4.35\times$ ideal speedup vs.\ $3.50\times$ for \name alone and $1.78\times$ for Eagle alone (Appendix~\ref{app:speedup_calc}).

\section{\name Runtime}\label{sec:design}

The \name runtime realizes the cross-resource staggering of \S\ref{sec:veri}: it decides which requests draft or verify at each iteration and schedules KV transfers so each verify lands on time. It maintains a resource model of the interconnect and GPU HBM, then places each request's next verify in a window with spare capacity---a placement made flexible by the high acceptance rate (\S\ref{subsec:p2}).

%If too many verifies fire in the same iteration, the interconnect congests, the GPU's compute spikes, and the GPU's HBM overflows---wiping out the throughput compression was meant to deliver (Figure~\ref{fig:phase_mix_illustrate}).
%The runtime must therefore stagger verify deadlines across a lookahead window so none of these limits is breached in any iteration; if no feasible deadline exists, the request falls back to a non-speculative path, specialized per setting in \S\ref{subsec:design_special}.

\subsection{Resource model}\label{subsec:design_model}

The runtime maintains $W$ future iterations horizon, indexed $i \in [0, W)$, tracking two reserve rings across the windows:

\paragraph{BW ring (interconnect).}
$T[i]$ holds the interconnect time already reserved for transfers landing in window $i$. 
The link is serialized---one transfer at a time, with a transfer's duration on the link equal to its size divided by the link's bandwidth $\text{BW}$---so the constraint is
\[
T[i] \le T_\text{iter} \quad \forall i \in [0, W).
\]
Both KV loads at request arrival (compressed for speculating requests, full for non-speculating) and verify reloads on each speculative round share this ring. For request $r$ with $\text{KV}_\text{full}^{(r)}$, the reload's iteration-equivalent duration is $\ell_r = \text{KV}_\text{full}^{(r)} / (\text{BW} \cdot T_\text{iter})$, spanning $S_r = \max(1,\,\lceil\ell_r\rceil)$ windows.

\paragraph{HBM ring (GPU memory).}
$B[i]$ holds the in-flight KV cache occupying HBM during window $i$---KV streaming onto the GPU for an upcoming verify but not yet consumed by it. Together with persistent residency, the constraint is
\[
M + \text{KV}_\text{resident} + B[i] \le \text{HBM} \quad \forall i \in [0, W),
\]
where $M$ is the model weights and $\text{KV}_\text{resident}$ counts every KV cache kept on the GPU between iterations---compressed for drafting requests, full for pinned or non-speculating ones.

GPU compute is not modeled as a third ring: \name's cross-resource staggering (\S\ref{subsec:p1}) spreads verifies evenly across iterations, so compute load is smoothed rather than bursting.

\subsection{Request admission and execution loop}\label{subsec:design_admit}

\paragraph{Request admission.}
\name runtime conducts an \textsc{Admit} operation (See Algorithm \ref{alg:admit}) at request arrival and again after every verify. 
It either picks a future verify iteration $d_r$ within its lookahead window---reserving the full-KV-cache transfer that must complete by $d_r$---or returns $r$ to waiting queue.

The search for the next verify iteration starts from checking whether the $x$-th iteration from now is feasible --- $x$ is the draft length that maximizes ideal speedup (Figure~\ref{fig:key_insight_2} in \S\ref{subsec:p2}), typically $20$--$50$. 
\name runtime looks at its resource model to see if there is sufficient interconnect to reload the full KV back to GPU likely before the iteration $x$ and whether there is sufficient HBM resource to hold the full KV. 
When $x$ does not work, the runtime checks the $x-1$th and the $x+1$th iteration from now, then $x-2$ and $x+2$, and so on; each candidate window indexes both rings simultaneously, and a reservation spans $S_r$ consecutive windows on both. 
The first candidate that satisfies the BW-ring and HBM-ring constraints (\S\ref{subsec:design_model}) is reserved; if none does, $r$ is returned to waiting queue.

\begin{algorithm}[t]
\caption{\textsc{Admit}$(r)$}
\label{alg:admit}
\begin{algorithmic}[1]
\State $\ell_r \gets \text{KV}_\text{full}^{(r)} / (\text{BW} \cdot T_\text{iter})$;\quad $S_r \gets \max(1,\,\lceil\ell_r\rceil)$
\State $\text{anchor} \gets \mathrm{clamp}(x,\; S_r,\; W{-}1)$
\State $\text{candidates} \gets \text{anchor}, \text{anchor}{\pm}1, \text{anchor}{\pm}2, \dots$ \Comment{clamped to $[S_r, W{-}1]$}
\ForAll{$d \in \text{candidates}$}
  \State $\text{span}_r \gets [d - S_r + 1,\; d]$
  \If{the BW-ring and HBM-ring constraints (\S\ref{subsec:design_model}) hold over $\text{span}_r$}
    \State reserve $r$ on $\text{span}_r$;\quad $d_r \gets d$;\quad $\text{mode}[r] \gets \textsc{Speculative}$
    \State \Return
  \EndIf
\EndFor
\State return $r$ to waiting queue
\end{algorithmic}
\end{algorithm}

\paragraph{Execution loop.}
At each iteration $t$, the runtime performs three steps:
(i)~\emph{Kick off verify reloads.}
For each speculating request whose reserved span has its first window at iteration $t$, start the asynchronous full-KV-cache reload on the link feeding the verifying GPU.
(ii)~\emph{Draft and verify.}
Draft run the next iteration's forward pass; concurrently, the verify completes the verify forward pass for each request whose scheduled iteration is the current one. 
For each completed verify, re-invoke \textsc{Admit}$(r)$ with the updated state---either continuing speculation with the next verify iteration scheduled, or returning $r$ to the waiting queue.
(iii)~\emph{Advance.}
Slide the lookahead window one iteration forward.

\subsection{Per-setting specialization}\label{subsec:design_special}

The two settings differ in their interconnects and in whether the verifying GPU is the same physical device as the drafter.

\paragraph{Long-context decoding.}
Drafting and verification share the serving GPU, and the interconnect is the CPU$\leftrightarrow$GPU link.
The BW ring (\S\ref{subsec:design_model}) tracks reservations on this link, and the HBM ring tracks the same GPU's HBM usage.
The compressed cache stays on GPU for drafting, and each verify reloads the full KV cache from CPU.

\paragraph{Remote prefix caching.}
Two GPU pools share a storage node (Figure~\ref{fig:setting}): a small local pool on a fast link $\text{BW}_h$, and a larger remote pool on a slow link $\text{BW}_l \ll \text{BW}_h$.
Remote-pool requests speculate: the remote GPU drafts using the compressed KV cache streamed over $\text{BW}_l$, while a local GPU concurrently loads the full KV cache over $\text{BW}_h$ for an upcoming verify and runs the verify forward pass for a request whose KV has already arrived. Because \name issues each load ahead of its verify deadline, drafting, loading, and verifying all pipeline together.
All links and pool HBMs are tracked by their own BW/HBM rings (\S\ref{subsec:design_model}): $\text{BW}_l$ carries initial compressed-KV loads to remote drafters, $\text{BW}_h$ carries verify reloads, and each pool's HBM ring tracks HBM usage.

% !TEX root = sample-sigconf.tex
\section{Compressor Interface}\label{sec:interface}

KV cache compression methods differ in what they produce: some drop token positions per (layer, head), others quantize entries to lower precision. \name's runtime relies on a small interface (Listing~\ref{lst:compressor}) that exposes only what it needs to allocate HBM and run attention---which positions are dropped, and bits per element.

\begin{lstlisting}[float=t,
                   caption={\name's compressor interface.},
                   breaklines=true,
                   label={lst:compressor}]
class CompressedKV:
    dropped_indices: list[Tensor]
    bit_scheme:      int         
class Compressor:
    scenario: Literal["long-context", "remote-prefix"]
    mode:     Literal["offline", "online"]
    def compress(full_kv, ratio) -> CompressedKV: ...
    def decompress(compressed, layer_idx=None, page_table=None): ...
    def update(layer_idx, q, k, v, hidden, req_offsets) -> list[Tensor]: ...
\end{lstlisting}

\begin{figure}[t]
    \centering
    \includegraphics[width=\columnwidth]{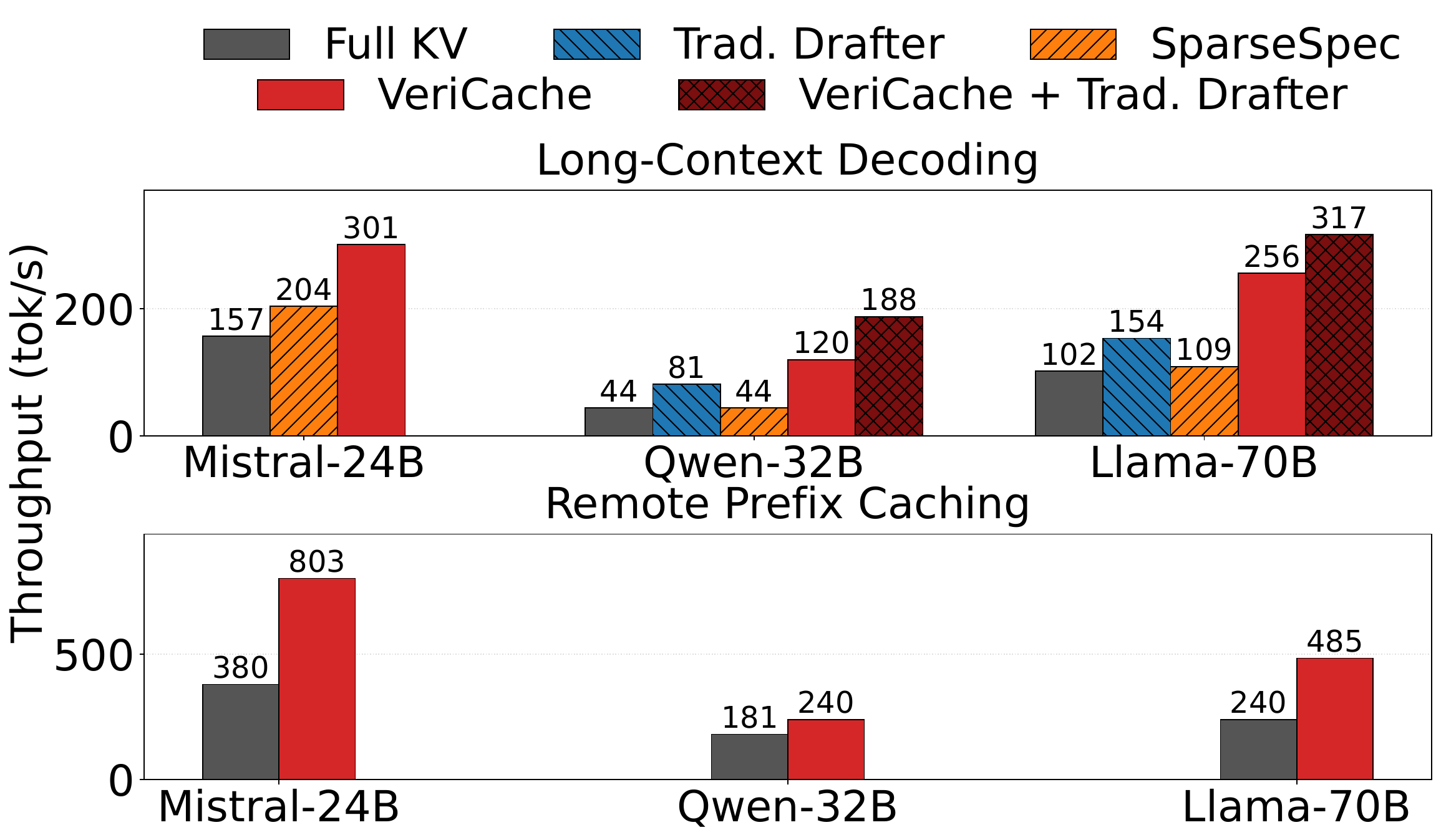}
    \caption{Sustained decoding throughput on long-context decoding and remote prefix caching.}
    \label{fig:e2e_speed_fullkv}
\end{figure}

\emph{Offline} means the compression step runs off the serving GPU (on CPU or an idle GPU), since methods often need information not computed by normal inference (e.g., explicit attention weights) or extra forward passes that would slow inline serving. \emph{Online} compression runs inline through \texttt{update}, with decisions driven by per-iteration state.

The interface is \emph{pass-through}: the runtime hands the compressor the physical-layout metadata and lets it gather, dequantize, or write pages itself. \texttt{update} mirrors this by passing \texttt{req\_offsets} for batch slicing. The cost---compressor authors must understand the runtime's paged layout---avoids a virtualization layer; KIVI's fused dequant-attention already operates on pages directly.

\paragraph{Long-context decoding.}
The runtime keeps the compressed cache in HBM and drafts against it. With offline compression, for example, KVzip~\cite{kvzip} scores tokens by context reconstructability and keeps, say, 25\% per (layer, head) of a 100K-token context; \texttt{compress} runs offline and returns \texttt{CompressedKV} with the 75K evicted positions per (layer, head) and \texttt{bit\_scheme=16}, after which the runtime allocates pages for the retained 25K and runs sparse attention with \texttt{decompress} never called. With online compression, for example, KVzap~\cite{kvzap} scores tokens during inference with a small MLP on hidden state: \texttt{compress} initializes \url{dropped\_indices} to empty; after each layer's forward, the runtime calls \url{update(layer, q, k, v, hidden, req\_offsets)} on the batched tensors, and the compressor slices the batch via \texttt{req\_offsets}, scores per request, and returns a \texttt{(num\_heads, new\_drops)} tensor per request. The runtime appends the new drops to \texttt{dropped\_indices[layer]} and trims the (layer, head) page allocation.

\paragraph{Remote prefix caching.}
The runtime never inspects the compressed cache: \texttt{compress} produces a payload at storage, the runtime streams the bytes, and \texttt{decompress(compressed)} materializes full KV at the remote drafter. The payload is a blob the compressor can fill however it wants.

\begin{figure*}[t]
    \centering
    \includegraphics[width=\textwidth]{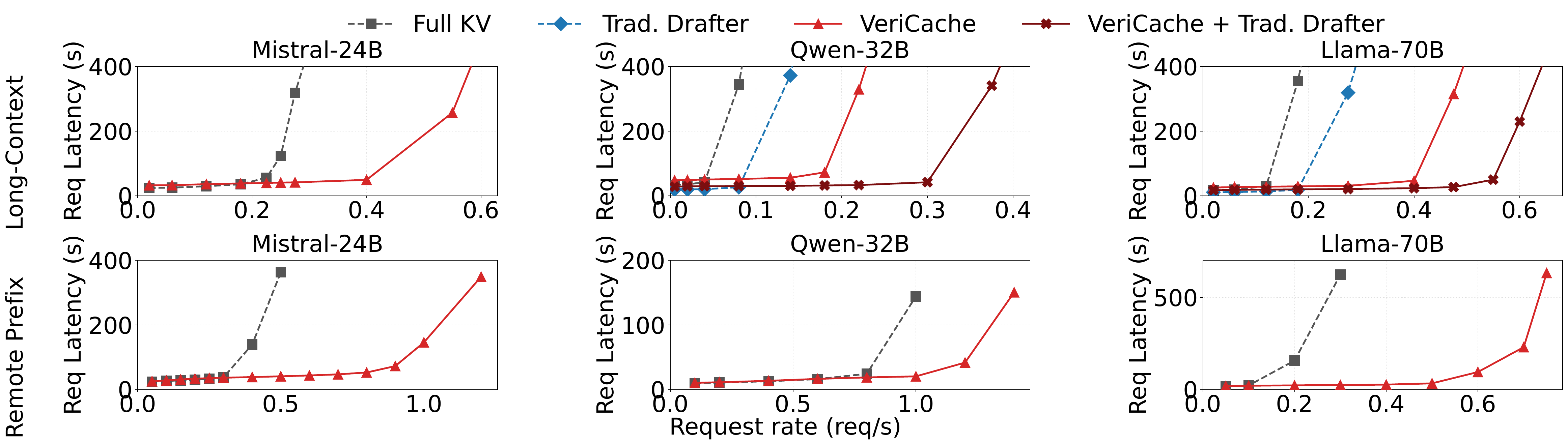}
    \caption{End-to-end request latency vs.\ request rate on Pipeline~1 (top) and Pipeline~2 (bottom).}
    \label{fig:eval_latency_throughput}
\end{figure*}

\paragraph{Scope.}
The interface targets \emph{deployment-time} substitution under four constraints:
(i)~heads within a layer may drop different positions but the same \emph{number} of tokens (count varies only across layers);
(ii)~the system runs one mode at a time; token-dropping and quantization do not mix across concurrent requests;
(iii)~\texttt{bit\_scheme} is uniform across tokens and layers;
(iv)~the compression method is fixed at deployment.
Each is a straightforward future-work extension. We instantiate the interface for seven methods.

\section{Implementation}\label{sec:impl}

\name is a thin layer between the request frontend and the GPU workers, built atop vLLM~\cite{vllm} (serving engine) and LMCache~\cite{lmcache} (persistent KV cache storage and transfer). \name is implemented with ~8K LoC of Python and C++.

\name subclasses vLLM's \texttt{AsyncScheduler} and manages its own GPU KV allocations so that compressed and transient reload KVs can coexist under our admission control. A scheduler hook runs \textsc{Admit} (Algorithm~\ref{alg:admit}) against the BW and HBM rings of \S\ref{subsec:design_model} and kicks each request's verify reload asynchronously $S_r$ windows ahead of its deadline; the bytes themselves move via LMCache, which exposes \texttt{lookup}/\texttt{lookup\_compressed} returning (compressed) KV pointers, \texttt{move} for cross-tier transfers (CPU$\leftrightarrow$GPU, storage$\leftrightarrow$GPU), and a wrapper that invokes the \texttt{Compressor} of \S\ref{sec:interface} on a stored KV pointer. Per-layer attention activations are routed through a forward hook to \texttt{Compressor.update} for online compressors; offline compressors run before serving (or on idle compute) and apply when the context's KV is reused.

% !TEX root = sample-sigconf.tex

\section{Results}\label{sec:eval}

\begin{figure*}[t]
    \centering
    \includegraphics[width=0.49\textwidth]{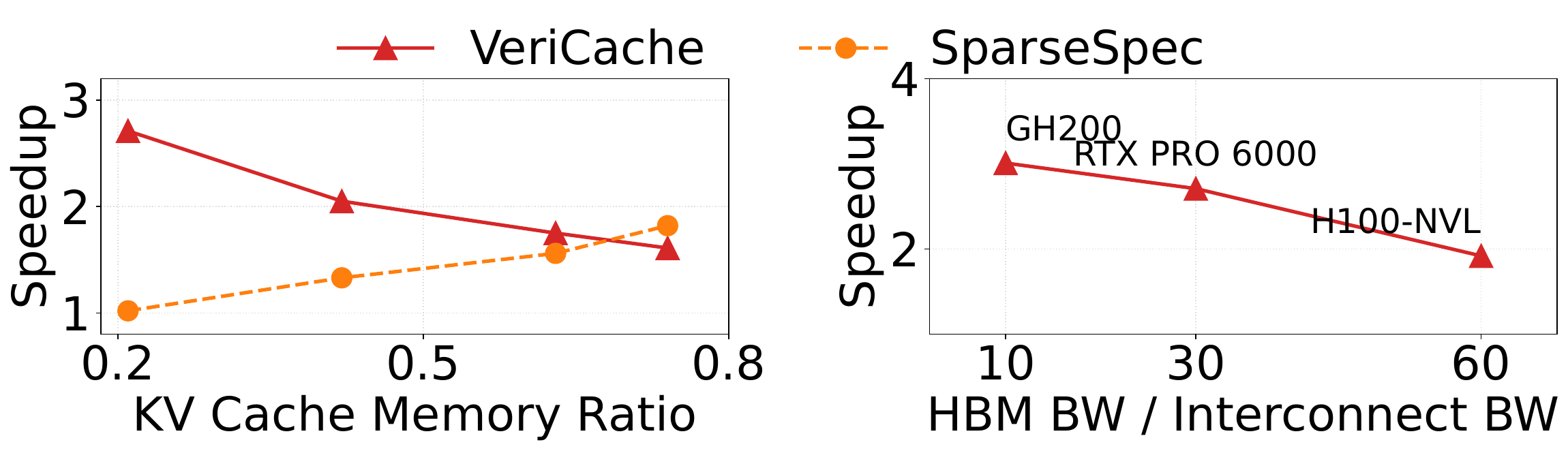}\hfill
    \includegraphics[width=0.49\textwidth]{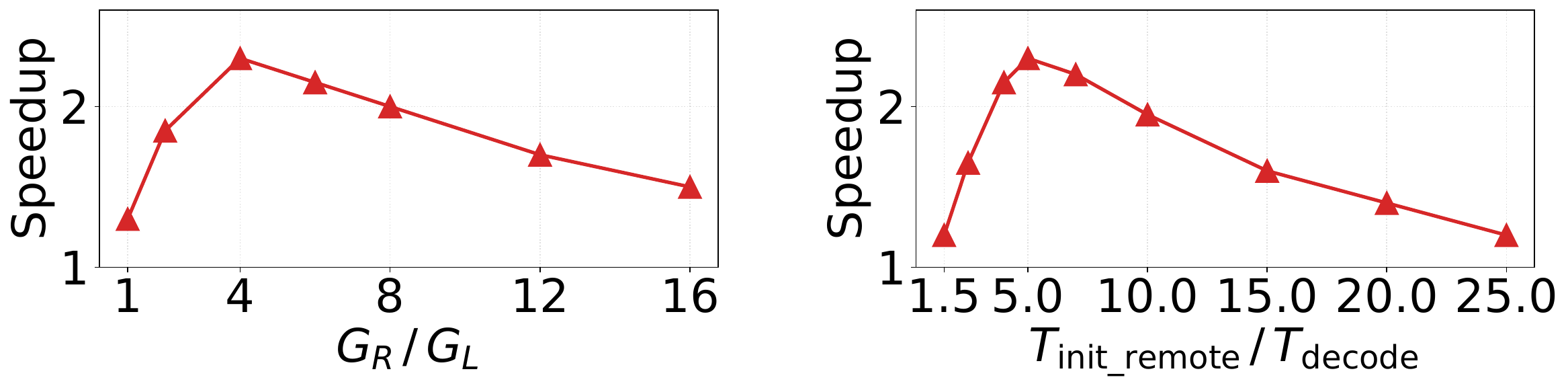}
    \caption{\name's speedup over Full~KV: Pipeline~1, varying KV-cache budget and HBM/interconnect ratio (left); Pipeline~2, varying $G_R/G_L$ and $T_{\mathrm{init\_remote}}/T_{\mathrm{decode}}$ (right).}
    \label{fig:eval_vary_hardware}
\end{figure*}

\subsection{Evaluation settings}\label{subsec:eval_settings}
\paragraph{Models and hardware.}
Mistral-24B~\cite{mistralsmall24b} and Qwen-32B~\cite{qwen3report} run on a NVIDIA RTX PRO 6000 (96\,GB); Llama-70B~\cite{grattafiori2024llama} runs on 2$\times$H100 NVL (94\,GB, TP$=$2). Each node connects CPU--GPU via PCIe 5.0 x16 (64\,GB/s); the local node reaches the KV store at $40$\,GB/s, remote nodes at $1.2$\,GB/s.

\paragraph{Two evaluation pipelines.}
Pipeline~1 (long-context decoding) precomputes the context's KV in CPU memory or storage and compresses it offline or online (\S\ref{sec:interface}); a single serving instance. Pipeline~2 (remote prefix caching) reuses KV across requests over the slow remote link with one local and four remote instances.

\paragraph{Compression methods.}
\emph{Token dropping}: KVzip~\cite{kvzip}, KVZap~\cite{kvzap}, ExpectedAttention~\cite{kvpress}, SnapKV~\cite{snapkv}. \emph{Quantization}: KIVI~\cite{kivi}, KVQuant~\cite{kvquant}, RotateKV~\cite{rotatekv}.

\paragraph{Speculative-decoding baselines.}
\emph{Traditional}: EAGLE3~\cite{eagle1, eagle2} auxiliary drafter, with RedHatAI pretrained speculators for Llama-70B~\cite{redhat_eagle3_llama70b} and Qwen-32B~\cite{redhat_eagle3_qwen32b}. \emph{Self-speculative}: SparseSpec~\cite{sparsespec}, which drafts on the target model with PillarAttn (sparse attention selecting critical tokens via prior-verification scores) and verifies with full attention over resident full KV.

\paragraph{Datasets and quality metrics.}
We use five datasets, each with its own quality metric:
\begin{itemize}
    \item \emph{LMCache-trace}~\cite{lmcache_trace}: both pipelines, offline+online (900 samples); KL-divergence from Full~KV.
    \item \emph{ComplexFuncBench}~\cite{complexfuncbench}: Pipeline~1 offline (150 samples); fraction of function calls with exactly-matching arguments.
    \item \emph{PISanitizer}~\cite{geng2025pisanitizer}: Pipeline~1 offline (50 samples); defense success rate against prompt injection.
    \item \emph{LongGenBench}~\cite{wu2024longgenbench}: Pipeline~1 online (50 samples); fraction of per-prompt constraints satisfied (completion rate).
    \item \emph{GSM8K-Long}~\cite{liu2024longgenbench}: Pipeline~1 online (100 samples); fraction of chained math problems answered exactly.
\end{itemize}

\paragraph{Efficiency metrics.} End-to-end \emph{request latency} (s) and \emph{throughput} (tok/s).

\begin{figure*}[t]
    \centering
    \includegraphics[width=\textwidth]{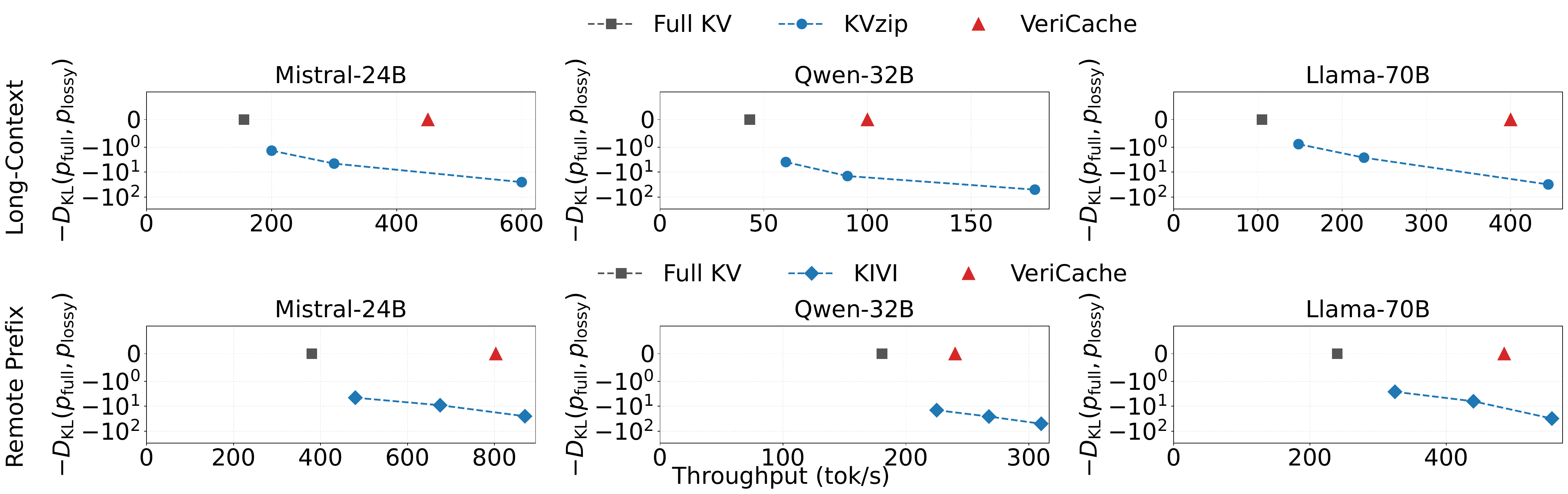}
    \caption{Quality (negative KL-divergence) vs.\ throughput on Pipeline~1 (top) and Pipeline~2 (bottom).}
    \label{fig:e2e_quality_speed}
\end{figure*}

\subsection{Comparing with full KV and traditional speculative decoding}\label{subsec:eval_vs_full}

Throughout this comparison, \name uses KVzip (compression ratio $0.2$, anchored draft length $x{=}25$) as its compressed-KV drafter on Pipeline~1 and KIVI (4-bit, $x{=}40$) on Pipeline~2.

Figure~\ref{fig:e2e_speed_fullkv} compares \name against Full~KV, a traditional drafter, SparseSpec, and \name~+~traditional drafter on three models. On long-context decoding, \name delivers $1.92\times$--$2.73\times$ over Full~KV (e.g., $256$ vs.\ $102$\,tok/s on Llama-70B), beating the traditional drafter, whose speedup is bounded by the unchanged KV footprint. Composing \name with the drafter peaks at $4.26\times$ (Qwen-32B), as the two attack orthogonal bottlenecks: \name shrinks per-request KV to enable larger batches, the drafter speeds up tokens within each request. On remote prefix caching, drafter-based methods do not apply; \name still gives $1.33\times$--$2.11\times$ over Full~KV (e.g., $485$ vs.\ $240$\,tok/s on Llama-70B).

\paragraph{Varying hardware (Pipeline~1).}
Figure~\ref{fig:eval_vary_hardware} sweeps the two hardware axes that dominate long-context decoding. Left: the KV-cache budget, varied by serving Qwen-8B, Qwen-14B, Mistral-24B, and Qwen-32B (larger weights leave less HBM for KV). As the budget shrinks from $0.74$ to $0.2$ of HBM, \name's speedup grows from $1.61\times$ to $2.71\times$---Full~KV's batch collapses faster than \name's, while SparseSpec drops from $1.82\times$ to $1.02\times$ since it must keep full KV resident on the drafting GPU. Right: as the HBM-to-interconnect ratio falls from $60$ (H100~NVL) to $10$ (GH200), \name's speedup rises from $1.92\times$ to $3.01\times$, as a faster interconnect makes each full-KV reload cheap, so verification fires more often without stalling drafting.

% \begin{figure}
%     \centering
%     \includegraphics[width=\columnwidth]{figs/e2e_quality_speed_extras_p1.pdf}\\
%     \includegraphics[width=\columnwidth]{figs/e2e_quality_speed_extras_p2.pdf}
%     \caption{Additional token-dropping and quantization baselines on Mistral-24B.}
%     \label{fig:e2e_quality_speed_extras}
% \end{figure}

\paragraph{Varying hardware (Pipeline~2).}
Figure~\ref{fig:eval_vary_hardware} (Left) sweeps the number of remote gpus divided by the number of local gpus ($G_R/G_L$), which sets whether the local and remote pools produce and consume speculated tokens at matching rates. Below the sweet spot at $G_R/G_L{=}4$, the remote pool under-supplies drafts and spare local cycles fall back to Full~KV; above it, local HBM cannot verify drafts fast enough and spare remote cycles fall back instead---in both regimes the speedup converges back toward $1\times$. Figure~\ref{fig:eval_vary_hardware} (Right) sweeps $T_{\mathrm{init\_remote}}/T_{\mathrm{decode}}$, the ratio of remote-KV arrival time to one draft-plus-verify round. Up to $5$, streaming a quantized KV over the slow remote network pays off since initial loading dominates the request lifecycle; beyond that, $T_{\mathrm{decode}}$ dominates and the benefit fades back toward $1\times$.

\subsection{Comparing with lossy KV}\label{subsec:eval_vs_compression}

\begin{figure}
    \centering
    \includegraphics[width=\columnwidth]{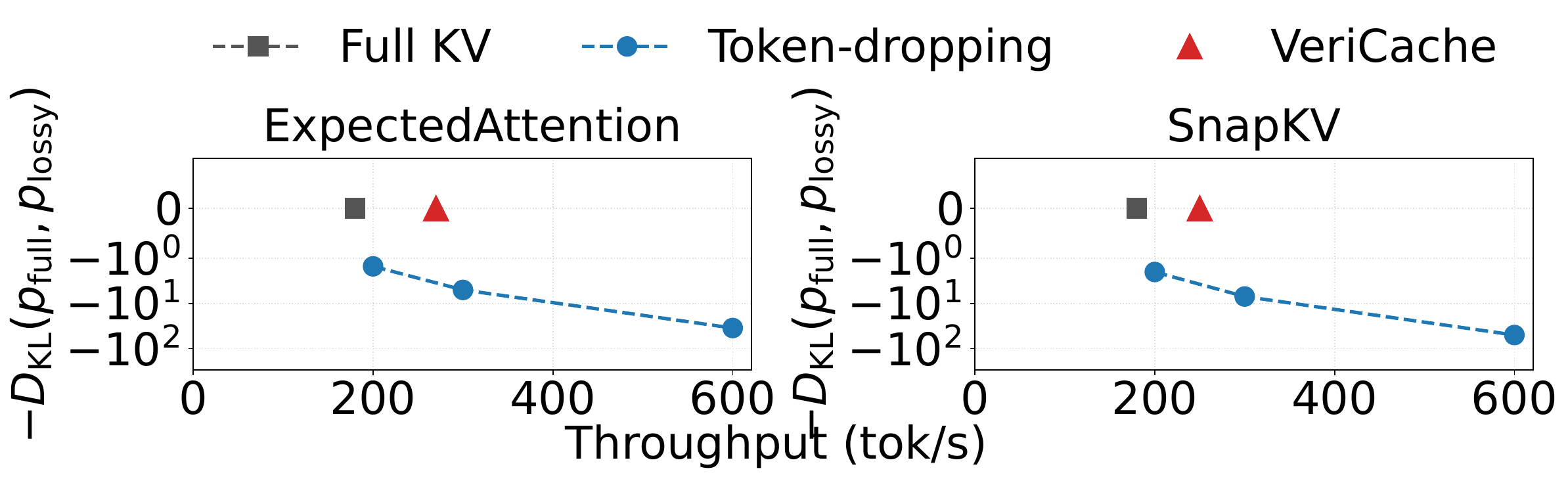}\\
    \includegraphics[width=\columnwidth]{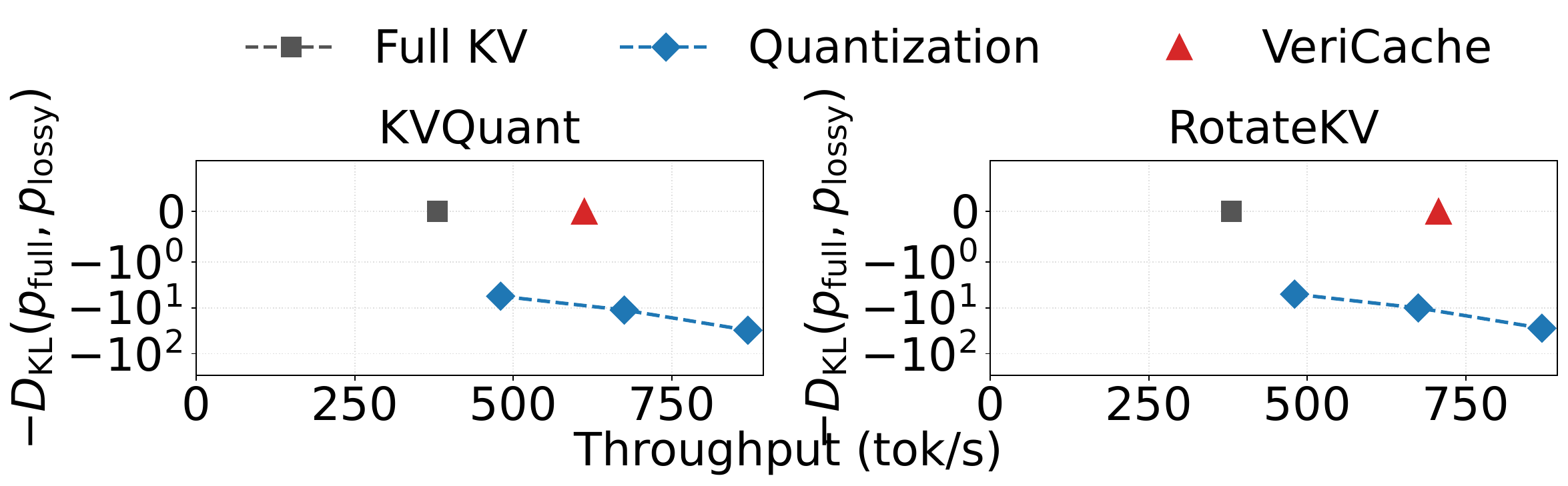}
    \caption{Additional token-dropping and quantization baselines on Mistral-24B.}
    \label{fig:e2e_quality_speed_extras}
\end{figure}

\begin{figure*}[!tbp]
    \centering
    \includegraphics[width=\textwidth]{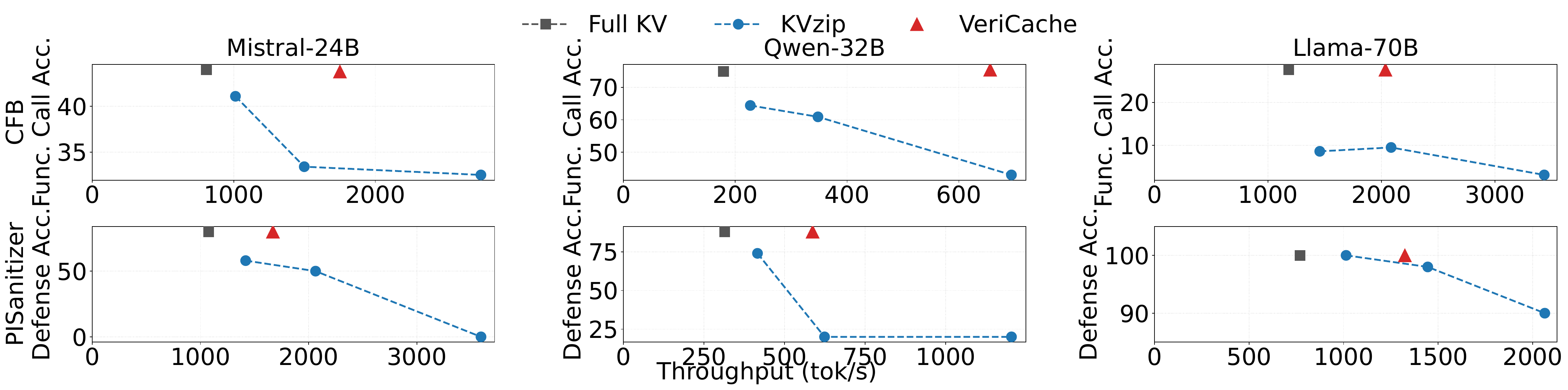}
    \caption{Quality vs.\ throughput: function-call accuracy (top), defense success rate (bottom).}
    \label{fig:e2e_quality_speed_cfb}
\end{figure*}

\begin{figure}[!htbp]
    \centering
    \includegraphics[width=\columnwidth]{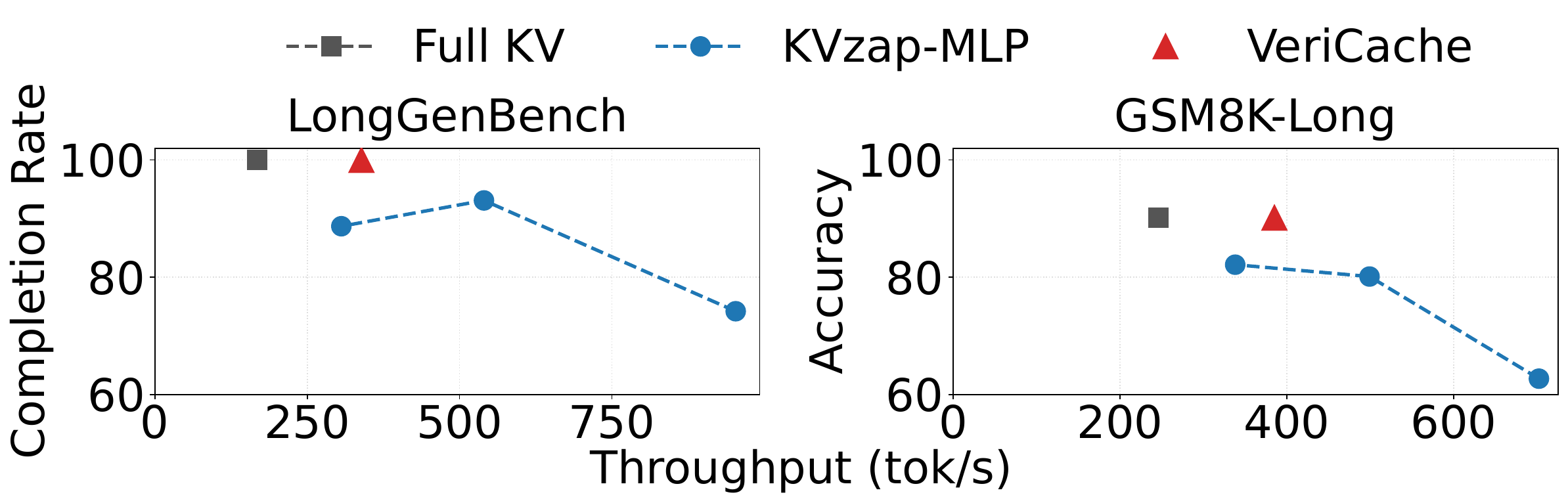}
    \caption{Quality vs.\ throughput on Pipeline~1 long-generation: completion rate (left), accuracy (right).}
    \label{fig:e2e_quality_speed_longgen}
\end{figure}

For each lossy baseline we sweep its compression knob to trace the quality--throughput frontier: offline token-dropping methods at compression ratios $0.25$, $0.5$, and $0.75$; online token-dropping methods at KV-cache budgets of $1024$, $2048$, and $4096$ tokens; and quantization methods at $8$, $4$, and $2$ bits.

\paragraph{Quality and throughput on two pipelines.}
Figure~\ref{fig:e2e_quality_speed} reports KL-divergence from Full~KV vs.\ decoding throughput across Pipeline~1 (top), Pipeline~2 (bottom), and three models. \name's KL stays under $0.01$ nats (attributable to hardware nondeterminism~\cite{nondeterminism_llm}) while reaching up to $3.82\times$ Full~KV's throughput (Llama-70B, Pipeline~1). The lossy baselines (KVzip, KIVI) must trade quality for throughput: on Llama-70B/Pipeline~1 at compression $0.5$, KVzip accumulates ${\sim}14.4$\,nats of KL per request---the lossy model emits Full~KV's exact output with probability only $e^{-14.4}{\approx}5{\times}10^{-7}$. At the highest compression this widens by another ${\sim}12\times$.

\paragraph{Varying datasets.}
The same picture holds across task-specific quality metrics on both Pipeline~1 workloads. Figure~\ref{fig:e2e_quality_speed_cfb} reports function-call accuracy on ComplexFuncBench~\cite{complexfuncbench}: \name reaches at least $59\%$ of the fastest KVzip configuration's throughput at Full~KV accuracy, while KVzip loses up to ${\sim}30$ points at the same throughput and collapses to ${\sim}31\%$ of Full~KV's accuracy on Llama-70B even at the most conservative ratio. Figure~\ref{fig:e2e_quality_speed_longgen} runs long-generation on Qwen-32B with LongGenBench~\cite{wu2024longgenbench} (completion rate) and GSM8K-Long~\cite{liu2024longgenbench} (accuracy): \name preserves Full~KV's $100\%$ completion at $339$\,tok/s and $90\%$ accuracy at $385$\,tok/s. KVzip's quality drops by around 10 points at the same speed.

\paragraph{Varying compression methods.}
Figure~\ref{fig:e2e_quality_speed_extras} adds two token-dropping baselines (ExpectedAttention~\cite{kvpress}, SnapKV~\cite{snapkv}) and two quantization baselines (KVQuant~\cite{kvquant}, RotateKV~\cite{rotatekv}) on Mistral-24B. Across all four, \name's KL stays within $0.01$ nats while running $1.4$--$1.9\times$ faster than Full~KV; the baselines trace the same quality--throughput frontier as the KVzip/KIVI headlines, accumulating tens of nats of KL.

% \jiayi{TODO: add a figure for one more long-context dataset (in addition to
% ComplexFuncBench shown in Figure~\ref{fig:e2e_quality_speed_cfb}), e.g.,
% another function-calling / agentic benchmark to demonstrate generality.}

\section{Related Work}\label{sec:related}

\paragraph{KV cache compression.}
Token-dropping methods (KVzip~\cite{kvzip}, KVzap~\cite{kvzap}) and quantization (KIVI~\cite{kivi}, KVQuant~\cite{kvquant}) shrink the KV cache but are inherently lossy (\S\ref{sec:motivation}). \name turns them lossless by drafting with the compressed cache and verifying against the full one.

\paragraph{Speculative decoding.}
Speculative decoding accelerates a target model with a cheaper proposer (EAGLE~\cite{eagle1, eagle2}, MTP~\cite{fastmtp}, n-gram~\cite{specngram}); these compose with \name. Closer to us, MagicDec~\cite{magicdec}, QuantSpec~\cite{quantspec}, and SparseSpec~\cite{sparsespec} draft from a sparse/compressed cache but pin the full KV in HBM, ignore remote prefix caching, and hard-wire one compressor---\name addresses all three.

\paragraph{Prefill--decode disaggregation.}
Serving systems disaggregate prefill and decode onto separate GPU pools (DistServe~\cite{distserve}, Splitwise~\cite{splitwise}, TetriInfer~\cite{tetriinfer}, Mooncake~\cite{mooncake}). Decode nodes hit exactly the HBM-bandwidth bottleneck \name targets, so \name slots directly into a decode pool.

\section{Limitations \& Future Work}\label{sec:limitations}

\paragraph{Memory overhead.}
\name keeps the full KV cache on CPU (or storage) in addition to the compressed cache on GPU, so it incurs more storage overhead.

\paragraph{Static draft length.}
\name uses a fixed draft length per workload; a per-request adaptive policy driven by early accept/reject outcomes would handle heterogeneous compressors and contexts more gracefully.

\paragraph{Drafter-specific compression.}
Existing compressors optimize direct-serving accuracy; a compressor designed to maximize acceptance length at large draft horizons---a different objective---could push \name's throughput further.

\paragraph{Verification beyond compression.}
Other lossy KV techniques besides compression---for instance, reusing precomputed KV across non-prefix chunks (CacheBlend~\cite{cacheblend})---also produce outputs that diverge from full-KV decoding. Whether a draft-then-verify approach could help here would be an interesting question to explore.

\section{Conclusion}\label{sec:conclusion}

Lossy KV compression speeds up long-context LLM serving while lowering output quality. \name restores lossless inference by drafting from the compressed cache and verifying against the full cache, with cross-resource staggering and long acceptance runs hiding the verification cost. It delivers up to $4\times$ higher throughput than full-KV decoding while producing identical outputs, across both token-dropping and quantization methods.

\clearpage

\bibliographystyle{ACM-Reference-Format}
\bibliography{sample-base}
\texttt{.    }
\texttt{    }
\texttt{    }
\texttt{    }
\texttt{    }
\texttt{    }
\texttt{    }
\texttt{    }
\texttt{    }
\texttt{    }
\texttt{    }
\texttt{    }
\texttt{    }
\texttt{    }
\texttt{    }
\texttt{    }
\texttt{    }
\texttt{    }
\texttt{    }
\texttt{    }
\texttt{    }
\texttt{    }
\texttt{    }
\texttt{    }
\texttt{    }
\texttt{    }
\texttt{    }
\texttt{    }
\texttt{    }
\texttt{    }
\texttt{    }
\texttt{    }
\texttt{    }
\texttt{    }
\texttt{    }
\texttt{    }
\texttt{    }
\texttt{    }
\texttt{    }
\texttt{    }
\texttt{    }
\texttt{    }
\texttt{    }
\texttt{    }
\texttt{    }
\texttt{    }
\texttt{    }
\texttt{    }
\texttt{    .}
% !TEX root = sample-sigconf.tex
\newpage
\newpage
\appendix

\section{Proof of the KL Chain Rule for Autoregressive Distributions}\label{app:kl_chain}

We prove Eq.~(\ref{eq:kl_chain}) from Section~\ref{sec:motivation}: for autoregressive distributions $p_\text{full}$ and $p_\text{lossy}$ over a sequence $x_{1:T}$, the sequence-level KL divergence decomposes additively as
\begin{equation*}
    \mathrm{KL}\!\left(p_\text{full}(x_{1:T})\,\|\,p_\text{lossy}(x_{1:T})\right)
    \;=\; \sum_{t=1}^{T} \mathbb{E}_{x_{<t}\sim p_\text{full}}\!\left[\mathrm{KL}_t\right],
\end{equation*}
where $\mathrm{KL}_t \;\triangleq\; \mathrm{KL}\!\left(p_\text{full}(\cdot\mid x_{<t}) \,\|\, p_\text{lossy}(\cdot\mid x_{<t})\right)$ is the per-step KL at decoding step $t$ given the prefix $x_{<t}$.

\paragraph{Proof.}
For brevity, let
\begin{equation*}
    \ell_t(x_{\le t}) \;\triangleq\; \log \frac{p_\text{full}(x_t \mid x_{<t})}{p_\text{lossy}(x_t \mid x_{<t})}
\end{equation*}
denote the per-step log-likelihood ratio.
By the chain rule of probability, both joint distributions factor as $p(x_{1:T}) = \prod_{t=1}^{T} p(x_t \mid x_{<t})$, so
\begin{equation*}
    \log \frac{p_\text{full}(x_{1:T})}{p_\text{lossy}(x_{1:T})}
    \;=\; \sum_{t=1}^{T} \ell_t(x_{\le t}).
\end{equation*}
By definition, the sequence-level KL is the expected log-likelihood ratio under $p_\text{full}$:
\begin{align*}
    \mathrm{KL}_{1:T}
    &= \mathbb{E}_{x_{1:T} \sim p_\text{full}}\!\left[\log \tfrac{p_\text{full}(x_{1:T})}{p_\text{lossy}(x_{1:T})}\right] \\
    &= \sum_{t=1}^{T} \mathbb{E}_{x_{1:T} \sim p_\text{full}}\!\left[\ell_t(x_{\le t})\right].
\end{align*}
Each $\ell_t$ depends only on $x_{\le t}$, so by the tower property,
\begin{align*}
    \mathbb{E}_{x_{1:T} \sim p_\text{full}}\!\left[\ell_t(x_{\le t})\right]
    &= \mathbb{E}_{x_{<t} \sim p_\text{full}}\!\left[\mathbb{E}_{x_t \sim p_\text{full}(\cdot\mid x_{<t})}\!\left[\ell_t(x_{\le t})\right]\right] \\
    &= \mathbb{E}_{x_{<t} \sim p_\text{full}}\!\left[\mathrm{KL}_t\right],
\end{align*}
where the last equality uses the fact that the inner expectation, $\mathbb{E}_{x_t \sim p_\text{full}(\cdot\mid x_{<t})}[\ell_t]$, is exactly the per-step KL conditioned on $x_{<t}$. Substituting back,
\begin{equation*}
    \mathrm{KL}_{1:T}
    \;=\; \sum_{t=1}^{T} \mathbb{E}_{x_{<t} \sim p_\text{full}}\!\left[\mathrm{KL}_t\right]. \qquad\blacksquare
\end{equation*}

\paragraph{Implication.}
If the per-step KL is bounded below by some $\varepsilon > 0$ on average---i.e., $\mathbb{E}_{x_{<t}\sim p_\text{full}}[\mathrm{KL}_t] \geq \varepsilon$ for all $t$---then $\mathrm{KL}_{1:T} \geq \varepsilon T$ grows at least linearly in the output length $T$. Because $\mathrm{KL}_{1:T} = \mathbb{E}_{x \sim p_\text{full}}[\log(p_\text{full}(x_{1:T})/p_\text{lossy}(x_{1:T}))]$, this means the expected log-likelihood ratio under $p_\text{full}$ grows linearly in $T$. Equivalently, for a sample $x_{1:T} \sim p_\text{full}$, the log-likelihood ratio $\log(p_\text{full}(x_{1:T})/p_\text{lossy}(x_{1:T}))$ has mean ${\geq}\, \varepsilon T$, so the likelihood ratio itself is typically of order $e^{\varepsilon T}$---i.e., exponential in $T$.

\section{Full Theoretical Analysis}\label{app:theoretical_analysis}

This appendix provides the complete derivations for the throughput models summarized in Section~\ref{sec:veri}.

\begin{table}[h]
\centering
\caption{Notation for theoretical analysis.}
\label{tab:notation_app}
\footnotesize
\setlength{\tabcolsep}{4pt}
\begin{tabular}{l l}
\toprule
\textbf{Symbol} & \textbf{Description} \\
\midrule
\multicolumn{2}{l}{\textit{Notations (shared)}} \\
\midrule
$M$ & Model weights size \\
$\text{KV}_\text{full}$ & Full KV cache size per request \\
$c$ & Compression fraction ($\text{KV}_\text{compressed} = c \cdot \text{KV}_\text{full}$, $c < 1$) \\
$\gamma(x,c)$ & Token acceptance rate (function of draft length and compression) \\
$x$ & Draft (speculation) length \\
$\text{BW}_\text{hbm}$ & GPU HBM bandwidth \\
$\text{GPU}_\text{mem}$ & GPU memory capacity \\
\midrule
\multicolumn{2}{l}{\textit{Notations (intra-request token dropping)}} \\
\midrule
$\text{BW}_\text{inter}$ & CPU--GPU interconnect bandwidth \\
$B$ & Total batch size (number of concurrent requests) \\
$B_g$ & Number of non-offloaded requests (full KV on GPU) \\
$B_c$ & Number of offloaded requests (full KV on CPU, $B_g + B_c = B$) \\
$\ell$ & Cycles per load (iterations to complete one full-KV reload) \\
\midrule
\multicolumn{2}{l}{\textit{Notations (inter-request KV quantization)}} \\
\midrule
$G_L$ & Number of local GPUs (close to storage) \\
$G_R$ & Number of remote GPUs (far from storage) \\
$\text{BW}_h$ & Storage--local GPU bandwidth (high) \\
$\text{BW}_l$ & Storage--remote GPU bandwidth (low) \\
$K$ & Number of output tokens per request \\
\bottomrule
\end{tabular}
\end{table}

\subsection{Intra-request KV access}\label{app:intra}

% \jiayi{Several more things not included in main tex (some of them have been included here but some are still not) (1) different rquests have different kv sizes; (2) verification request should be drafting request divided by x+1; (3) we should use floor division; (4) need to differentiate offloaded reqeusts and non-offloaded requests; (5) for some methods we can load (1-c)kv whilw for some we need to load kv.}

We consider a single GPU serving $B$ homogeneous long-context decoding requests in steady state, where each request has the same context length and thus the same KV cache size $\text{KV}_\text{full}$. We focus on the decode phase, which is HBM-bandwidth-bound for long-context workloads.
Table~\ref{tab:notation_app} summarizes the notation used throughout this analysis.
All requests speculate: each cycles through $x$ draft iterations (using compressed KV) followed by one verification iteration (using full KV).
A request is either \emph{non-offloaded} ($B_g$: full KV remains on GPU) or \emph{offloaded} ($B_c$: full KV resides on CPU).
Non-offloaded requests verify for free since the full KV is already on GPU; offloaded requests must reload the full KV from CPU for verification.
In steady state, $\frac{x}{x+1}B_c$ offloaded requests are drafting and $\frac{1}{x+1}B_c$ are verifying, giving an average GPU KV footprint per offloaded request of:
\begin{equation}\label{eq:kv_avg_app}
    \text{KV}_\text{avg} = \text{KV}_\text{full} \cdot \frac{xc + 1}{x + 1}.
\end{equation}

\textit{GPU memory constraint.}
Since $\text{KV}_\text{avg} < \text{KV}_\text{full}$, \name serves a larger batch than the baseline:
\begin{equation}\label{eq:mem_constraint_app}
    M + B_g \cdot \text{KV}_\text{full} + B_c \cdot \text{KV}_\text{avg} \leq \text{GPU}_\text{mem}.
\end{equation}

\textit{Per-iteration time.}
The GPU time is bounded by HBM bandwidth:
\begin{equation}\label{eq:t_gpu_app}
    T_\text{gpu} = \frac{M + B \cdot \text{KV}_\text{avg}}{\text{BW}_\text{hbm}}.
\end{equation}
Each offloaded verification requires transferring the extra KV of size $(1{-}c) \cdot \text{KV}_\text{full}$ from CPU to GPU.
Rather than assuming each reload completes within a single iteration, we let $\ell$ denote the number of iterations a single reload spans (cycles per load).
Spreading the transfer over $\ell$ iterations reduces the per-iteration interconnect demand:
\begin{equation}\label{eq:t_xfer_app}
    T_\text{xfer} = \frac{B_c \cdot (1-c) \cdot \text{KV}_\text{full}}{(x+1) \cdot \text{BW}_\text{inter} \cdot \ell}.
\end{equation}
However, the interconnect can only sustain one in-flight reload at a time, which constrains the arrival rate of new loads:
\begin{equation}\label{eq:load_constraint_app}
    \frac{B_c}{x+1} \cdot \ell \leq 1.
\end{equation}
Since HBM reads and interconnect transfers use distinct hardware, they overlap:
\begin{equation}\label{eq:t_iter_app}
    T_\text{iter} = \max(T_\text{gpu},\; T_\text{xfer}).
\end{equation}

\textit{Throughput.}
All requests produce $\frac{\gamma(x,c) \cdot x + 1}{x+1}$ tokens per iteration on average, where $\gamma(x,c)$ is the acceptance rate as a function of draft length and compression ratio.
The scheduler chooses $B_c$, $x$, $c$, and $\ell$ to maximize throughput:
\begin{align}\label{eq:opt_app}
    \max_{B_c,\, x,\, c,\, \ell} \;\; &\frac{B \cdot \frac{\gamma(x,c) \cdot x + 1}{x+1}}{\max\!\left(T_\text{gpu},\;\; T_\text{xfer}\right)} \nonumber \\
    \text{s.t.} \;\; &\text{Eqs.~(\ref{eq:mem_constraint_app}),~(\ref{eq:load_constraint_app}),}\; 0 \leq B_c \leq B, \nonumber \\
    &x \geq 1,\; 0 < c < 1,\; \ell \geq 1.
\end{align}
The four knobs navigate a trade-off between $T_\text{gpu}$, $T_\text{xfer}$, and the effective tokens per iteration. Increasing $B_c$ lowers $T_\text{gpu}$ (smaller per-request HBM footprint enables larger batches and faster reads) but raises $T_\text{xfer}$ (more offloaded requests require more interconnect transfers). Increasing $\ell$ reduces $T_\text{xfer}$ by spreading each reload over more iterations, but the constraint $\frac{B_c}{x+1} \cdot \ell \leq 1$ forces either fewer offloaded requests or longer draft sequences. Decreasing $c$ further reduces $T_\text{gpu}$ by shrinking the compressed KV size, but also lowers $\gamma(x,c)$, reducing the number of accepted tokens per iteration. Increasing $x$ amortizes the interconnect cost over more draft tokens (lowering $T_\text{xfer}$), but $\gamma(x,c)$ decreases with longer drafts, yielding diminishing returns in accepted tokens.

The full KV baseline ($B_c{=}0$) serves at most $\lfloor(\text{GPU}_\text{mem} {-} M) / \text{KV}_\text{full}\rfloor$ requests;
pure lossy decoding serves up to $1/c\times$ more but sacrifices exactness.
\name operates between these extremes while maintaining lossless output.

\subsection{Inter-request KV reuse}\label{app:inter}

% \jiayi{Several more things not included in main tex (some of them have been included here but some are still not): (1) stateless vs cached (not all paths are modeled); (2) only one resource constraint is modeled in main tex; (3) heterogeneous gpus/cpus and links; (4) different output lengths; (5) different requests have different kv sizes; (6) we should use floor division; (7) need to differentiate offloaded reqeusts and non-offloaded requests; (8) for some methods we can load (1-c)kv while for some we need to load kv; (9) memory usage should be correctly monitored}

We consider a cluster with $G_L$ local GPUs close to a KV storage node (bandwidth $\text{BW}_h$) and $G_R$ remote GPUs connected via a slower link (bandwidth $\text{BW}_l$, with $\text{BW}_h \gg \text{BW}_l$).
Each request reuses a pre-computed KV cache from storage and must generate $K$ output tokens. Here, for simplicity, we assume all requests have the same KV size and output length.
We consider KV quantization methods where the KV cache is compressed offline to reduce network transfer size ($c \cdot \text{KV}_\text{full}$ bytes) but is decompressed to full size on the GPU upon reuse.

\textit{Derived quantities.}
Each GPU can batch at most $B_\text{max} = \lfloor(\text{GPU}_\text{mem} {-} M) / \text{KV}_\text{full}\rfloor$ requests.
The per-request per-token decode time depends on the actual batch occupancy $B$ of that GPU:
\begin{equation}\label{eq:t_tok_app}
    T_\text{tok}(B) = \frac{M / B + \text{KV}_\text{full}}{\text{BW}_\text{hbm}}.
\end{equation}
Since model weights $M$ are shared across the batch, $T_\text{tok}$ decreases as $B$ grows and reaches its minimum at $B = B_\text{max}$.
Network transfer times per request are $T_h = \text{KV}_\text{full} / \text{BW}_h$ (full KV to local), $T_l = \text{KV}_\text{full} / \text{BW}_l$ (full KV to remote), $T_l^c = c \cdot \text{KV}_\text{full} / \text{BW}_l$ (compressed KV to remote), and $T_l^\text{rem} = (1{-}c) \cdot \text{KV}_\text{full} / \text{BW}_l$ (remaining KV to remote).

\textit{Resource model.}
We model each GPU pool with three resources: \emph{network time} (how long the per-GPU network link is active), \emph{GPU compute time} (how long the GPU computes), and \emph{memory-time} (how long a memory slot is occupied).
Although network and compute are serialized within a single request, they use distinct hardware and can be \emph{pipelined across requests}: one request's KV transfer overlaps with another request's decode on the same GPU.
The capacity constraints are:
\begin{align}\label{eq:cap_app}
    \textstyle\sum_i L_i^{\text{net}} \cdot n_i &\leq G_L, \\
    \textstyle\sum_i L_i^{\text{gpu}} \cdot n_i &\leq G_L, \\
    \textstyle\sum_i L_i^{\text{mem}} \cdot n_i &\leq G_L \cdot B_\text{max}, \\
    \textstyle\sum_i R_i^{\text{net}} \cdot n_i &\leq G_R, \\
    \textstyle\sum_i R_i^{\text{gpu}} \cdot n_i &\leq G_R, \\
    \textstyle\sum_i R_i^{\text{mem}} \cdot n_i &\leq G_R \cdot B_\text{max},
\end{align}
where $n_i$ is the throughput (requests per second) assigned to path $i$, and $L_i^{\text{net}}$, $L_i^{\text{gpu}}$, $L_i^{\text{mem}}$, $R_i^{\text{net}}$, $R_i^{\text{gpu}}$, $R_i^{\text{mem}}$ are the per-request local and remote costs (in seconds).
Memory-time equals the wall-clock duration a slot is occupied (transfer plus compute), and may exceed GPU compute time when KV is cached between verification rounds.

\textit{Baseline paths.}
Two paths are available without \name:
\begin{itemize}[leftmargin=*,itemsep=2pt]
    \item \textbf{B1 (pure local):} Transfer full KV to local, decode $K$ tokens. $L^{\text{net}} {=} T_h$, $L^{\text{gpu}} {=} K \cdot T_\text{tok}$, $L^{\text{mem}} {=} T_h + K \cdot T_\text{tok}$, $R {=} 0$.
    \item \textbf{B2 (pure remote):} Transfer full KV to remote, decode $K$ tokens. $R^{\text{net}} {=} T_l$, $R^{\text{gpu}} {=} K \cdot T_\text{tok}$, $R^{\text{mem}} {=} T_l + K \cdot T_\text{tok}$, $L {=} 0$.
\end{itemize}

\textit{\name paths.}
\name adds two speculative paths that exploit the gap between compressed and full transfer times.

\textbf{P1 (verify-only, no full KV to remote).}
Remote receives compressed KV ($T_l^c$) and drafts $K/\gamma$ tokens with lossy KV.
Local receives full KV ($T_h$) and verifies in $N_1 = \lceil K / (\gamma x) \rceil$ rounds of $x$ draft tokens each.
Since local GPU memory is limited, KV may be evicted between rounds. We model two variants:
\begin{itemize}[leftmargin=*,itemsep=2pt]
    \item \emph{Cached:} KV stays on local GPU across rounds. $L^{\text{net}} = T_h$, $L^{\text{gpu}} = N_1 \cdot T_\text{tok}$, $L^{\text{mem}} = T_h + (K/\gamma) \cdot T_\text{tok}$ (slot held during entire draft phase).
    \item \emph{Stateless:} KV re-transferred each round. $L^{\text{net}} = N_1 \cdot T_h$, $L^{\text{gpu}} = N_1 \cdot T_\text{tok}$, $L^{\text{mem}} = N_1 \cdot (T_h + T_\text{tok})$.
\end{itemize}
In both cases, $R^{\text{net}} = T_l^c$, $R^{\text{gpu}} = (K / \gamma) \cdot T_\text{tok}$, $R^{\text{mem}} = T_l^c + (K / \gamma) \cdot T_\text{tok}$.
P1 is efficient for short outputs where streaming full KV to remote is wasteful.

\textbf{P2 (stream full KV to remote + AR).}
Remote receives compressed KV ($T_l^c$) and begins drafting immediately.
In parallel, the remaining KV streams to remote ($T_l^\text{rem}$).
During this streaming window, remote produces $N_\text{draft} = \min(T_l^\text{rem} / T_\text{tok},\; K / \gamma)$ draft tokens, verified by local in $N_2 = \lceil N_\text{draft} / x \rceil$ rounds.
Once the full KV arrives, remote completes the remaining $K_2 = \max(0,\; K - \gamma \cdot N_\text{draft})$ tokens via standard decode.
\begin{itemize}[leftmargin=*,itemsep=2pt]
    \item Remote: $R^{\text{net}} = T_l^c + T_l^\text{rem}$, $R^{\text{gpu}} = (N_\text{draft} + K_2) \cdot T_\text{tok}$, $R^{\text{mem}} = T_l^c + \max(N_\text{draft} \cdot T_\text{tok},\; T_l^\text{rem}) + K_2 \cdot T_\text{tok}$.
    \item Local cached: $L^{\text{net}} = T_h$, $L^{\text{gpu}} = N_2 \cdot T_\text{tok}$, $L^{\text{mem}} = T_h + N_\text{draft} \cdot T_\text{tok}$.
    \item Local stateless: $L^{\text{net}} = N_2 \cdot T_h$, $L^{\text{gpu}} = N_2 \cdot T_\text{tok}$, $L^{\text{mem}} = N_2 \cdot (T_h + T_\text{tok})$.
\end{itemize}
P2 is efficient for long outputs where the streaming window amortizes the cost of transferring full KV.

\textit{Throughput optimization.}
With all six paths \{B1, B2, P1-cached, P1-stateless, P2-cached, P2-stateless\}, the system maximizes total throughput (in tok/s):
\begin{equation}\label{eq:opt_inter_app}
    \max_{\{n_i \geq 0\}} \;\; K \cdot \textstyle\sum_i n_i
    \quad \text{s.t. Eqs.~(\ref{eq:cap_app})}
\end{equation}
where the resource costs in Eqs.~(\ref{eq:cap_app}) are evaluated at the actual batch occupancies implied by $\{n_i\}$.
The optimizer naturally selects the best mix: P1 for short outputs, P2 for long outputs, cached verification when local memory is abundant, stateless when local compute is abundant, and baseline paths when speculation offers no benefit.

% \section{temporary pipeline2 pseudo-code}
%
% \begin{algorithm}[t]
% \caption{\textsc{Admit}$_{\text{P2}}(r)$}
% \label{alg:admit_p2}
% \begin{algorithmic}[1]
% \State $\text{ETA}_L \gets \max(t,\,b_L) + \text{KV}_\text{full}^{(r)}/\text{BW}_h$;\quad $\text{ETA}_R \gets \max(t,\,b_R) + c\,\text{KV}_\text{full}^{(r)}/\text{BW}_l$
% \If{$\text{ETA}_L \le \text{ETA}_R$}
%   \State reserve $L$ for $\text{KV}_\text{full}^{(r)}/\text{BW}_h$;\quad $\text{mode}[r] \gets \textsc{B1}$;\quad \Return
% \EndIf
% \State reserve $R$ for $c\,\text{KV}_\text{full}^{(r)}/\text{BW}_l$ \Comment{compressed init: always needed}
% \State $\ell_r \gets \text{KV}_\text{full}^{(r)} / (\text{BW}_h \cdot T_\text{iter})$;\quad $S_r \gets \max(1,\,\lceil\ell_r\rceil)$
% \State $\text{anchor} \gets \mathrm{clamp}(x,\; S_r,\; W{-}1)$
% \State $\text{candidates} \gets \text{anchor}, \text{anchor}{\pm}1, \text{anchor}{\pm}2, \dots$ \Comment{clamped to $[S_r, W{-}1]$}
% \ForAll{$d \in \text{candidates}$}
%   \State $\text{span}_r \gets [d - S_r + 1,\; d]$
%   \If{the BW$_h$-ring and HBM-ring constraints (\S\ref{subsec:design_model}) hold over $\text{span}_r$}
%     \State reserve $r$ on $\text{span}_r$;\quad schedule bg $(1{-}c)\text{KV}_\text{full}^{(r)}$ on $R$
%     \State $d_r \gets d$;\quad $\text{mode}[r] \gets \textsc{Speculative}$;\quad \Return
%   \EndIf
% \EndFor
% \State reserve $R$ for $(1{-}c)\text{KV}_\text{full}^{(r)}/\text{BW}_l$;\quad $\text{mode}[r] \gets \textsc{B2}$ \Comment{demote (\S\ref{subsec:design_special})}
% \end{algorithmic}
% \end{algorithm}

\section{Ideal Speedup Calculation for Speculative-Decoding Comparisons}\label{app:speedup_calc}

% \jiayi{describe how the ideal speedup numbers in Figures~\ref{fig:benefit1} and~\ref{fig:benefit3} are computed for traditional speculative decoding baselines (Eagle, small-model drafter): assumed hardware config, batch size, KV size, draft model overhead, etc., and the formula used.}

This calculation extends the intra-request throughput model of
Appendix~\ref{app:intra} to a two-level speculation scheme.
\name proposes $x$ outer tokens using the target model on compressed KV
(the same $x$ and $\gamma(x,c)$ as Appendix~\ref{app:intra}); an auxiliary
drafter (Eagle, MTP-head) proposes $d_e - 1$ additional tokens at each
outer position.

Under chain-independence of the two drafters, the expected accepted draft
length per cycle becomes
\begin{equation*}
\gamma(x, c)\cdot x \cdot \big[\,1 + \gamma_e(d_e)\cdot(d_e - 1)\,\big],
\end{equation*}
i.e., \name's accepted length $\gamma(x,c)\cdot x$ from
Eq.~(\ref{eq:opt_app}) scaled by the composition multiplier
$[1 + \gamma_e(d_e)(d_e-1)]$.

% \section{Remote Prefix Caching Request Flow}\label{app:impl_remote}
%
% \jiayi{add request flow for remote}

\end{document}